\documentclass[sigconf]{acmart}

\AtBeginDocument{%
  \providecommand\BibTeX{{%
    \normalfont B\kern-0.5em{\scshape i\kern-0.25em b}\kern-0.8em\TeX}}}

\setcopyright{acmcopyright}
\copyrightyear{2023}
\acmYear{2023} 
\setcopyright{rightsretained}
\acmConference[CHI '23]{Proceedings of the 2023 CHI Conference on Human Factors in Computing Systems}{April 23--28, 2023}{Hamburg, Germany}
\acmBooktitle{Proceedings of the 2023 CHI Conference on Human Factors in Computing Systems (CHI '23), April 23--28, 2023, Hamburg, Germany}
\acmDOI{10.1145/3544548.3581236} 
\acmISBN{978-1-4503-9421-5/23/04}

\usepackage{caption}
\usepackage{subcaption}
\usepackage{enumitem}
\usepackage{soul}
\usepackage{acmart-taps}

\setstcolor{black}

\newcommand{\bc}[1]{\textcolor{black}{#1}}

\newcommand{\ehud}[1]{\textcolor{black}{#1}}
\newcommand{\add}[1]{\textcolor{black}{#1}}

\usepackage{newverbs}
\newverbcommand{\variable}{\color{blue}}{}
\newverbcommand{\function}{\color{purple}}{}
\newverbcommand{\argument}{\color{teal}}{}

\def\function#1{\textcolor{purple}{#1}}
\def\argument#1{\textcolor{teal}{#1}}
\def\variable#1{\textcolor{blue}{#1}}

\usepackage{tikz}

\begin{document}

\title{Causalvis: Visualizations for Causal Inference}

\author{Grace Guo}
\orcid{0000-0001-8733-6268}
\affiliation{%
  \institution{Georgia Institute of Technology}
  \streetaddress{North Ave NW}
  \city{Atlanta}
  \state{Georgia}
  \country{USA}
  \postcode{30332}
}
\email{gguo31@gatech.edu}

\author{Ehud Karavani}
\orcid{0000-0002-0187-5437}
\affiliation{%
  \institution{IBM Research}
  \streetaddress{}
  \city{Givatayim}
  \state{Tel Aviv}
  \country{Israel}
  \postcode{3498825}
}
\email{ehudk@ibm.com}

\author{Alex Endert}
\orcid{0000-0002-6914-610X}
\affiliation{%
  \institution{Georgia Institute of Technology}
  \streetaddress{North Ave NW}
  \city{Atlanta}
  \state{Georgia}
  \country{USA}
  \postcode{30332}
}
\email{endert@gatech.edu}

\author{Bum Chul Kwon}
\orcid{0000-0002-9391-6274}
\affiliation{%
  \institution{IBM Research}
  \streetaddress{314 Main St.}
  \city{Cambridge}
  \state{Massachusetts}
  \country{USA}
  \postcode{02138}
}
\email{bumchul.kwon@us.ibm.com}


\begin{abstract}
  Causal inference is a statistical paradigm for quantifying causal effects using observational data.
  It is a complex process, requiring multiple steps, iterations, and collaborations with domain experts.
  Analysts often rely on visualizations to evaluate the accuracy of each step.
  However, existing visualization toolkits are not designed to support the entire causal inference process within computational environments familiar to analysts.
  In this paper, we address this gap with Causalvis, a Python visualization package for causal inference.
  Working closely with causal inference experts, we adopted an iterative design process to develop four interactive visualization modules to support causal inference analysis tasks.
  The modules are then presented back to the experts for feedback and evaluation.
  We found that Causalvis effectively supported the iterative causal inference process.
  We discuss the implications of our findings for designing visualizations for causal inference, particularly for tasks of communication and collaboration.
\end{abstract}

\ccsdesc[300]{Human-centered computing~Visualization design and evaluation methods}
\ccsdesc[500]{Human-centered computing~Visualization toolkits}
\ccsdesc[500]{Human-centered computing~Visual analytics}

\keywords{causal inference, causality, design study}

\begin{teaserfigure}
  \includegraphics[width=\textwidth]{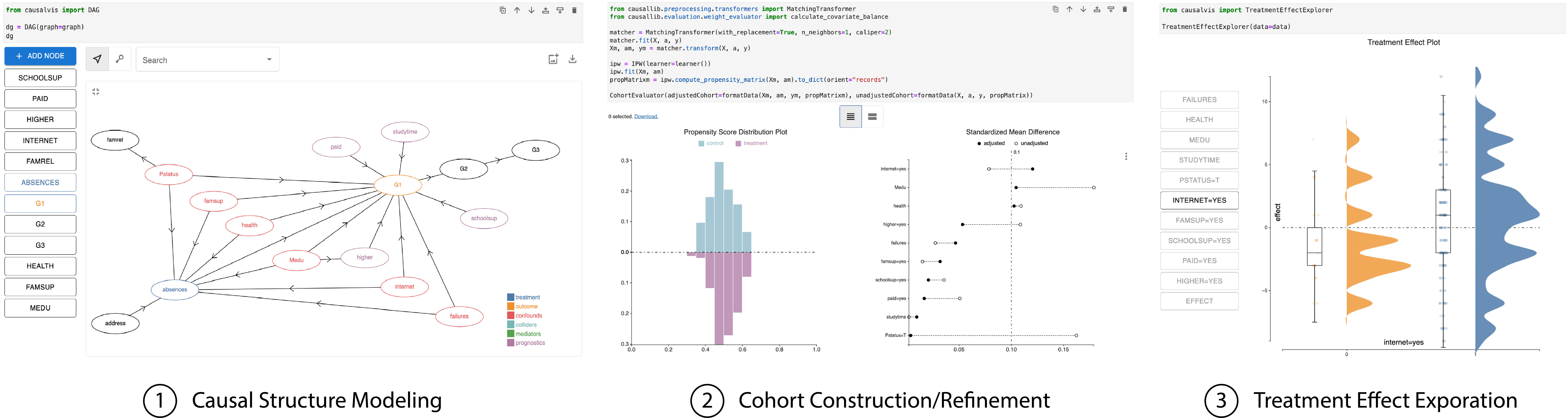}
  \caption{The Causalvis visualization modules corresponding to the three steps of the causal inference workflow: 1) Causal Structure Modeling, 2) Cohort Construction/Refinement, and 3) Treatment Effect Exploration.}
  \label{fig:teaser}
  \Description{Three screenshots of the Causalvis visualization modules from left to right. Below each screenshot is a numbered label naming the corresponding step in the workflow.}
\end{teaserfigure}

\maketitle

\section{Introduction}

\textit{Causal inference} is a statistical paradigm for quantifying causal effects using observational data.
Although randomized controlled trials (RCTs) are considered the gold standard for establishing causality, they are not always feasible, especially in situations where manipulating the treatment variable is costly, impractical, or unethical.
In medical fields, for instance, it would be highly unethical to randomly select a group of participants and ask them to smoke cigarettes for a period of time in order to measure the effect of smoking on lung cancer.
Causal inference, which requires only observational data, is thus a powerful approach that has been widely applied in healthcare \cite{glass2013causal, hernan2021methods}, economics \cite{athey2017state}, and the social sciences \cite{gerring2005causation}.
Researchers interested in studying the effect of a particular treatment exposure can select two groups from an observational data set -- treatment and control -- and compare outcomes between them.
Estimating the size of the effect of treatment exposure on the outcome is often the goal of causal inference analysis.

By replacing RCTs with observational data, additional steps must be taken to ensure that the estimated treatment effect is reliable and unbiased.
This is a complex process, requiring multiple steps, iterations, and collaborations with domain experts.
Researchers need to accurately understand the causal relationships in the data set,
identify and control for all confounding variables, make the necessary statistical adjustments, and ensure that the selected treatment and control groups satisfy required assumptions.
Consider again the scenario where researchers are interested in estimating the effect of cigarettes on lung cancer.
Researchers can begin by selecting two groups -- smokers and non-smokers -- from a large data set of medical records, then compare rates of lung cancer (outcome) between those that smoke (treatment) versus those that do not (control).
However, for this comparison to be valid, several assumptions must hold true. One such assumption is that the groups must share some minimal amount of similarities -- for example, if we have no women in the smoking group, we should not include women in the control group either. Moreover, there should be no unmeasured or uncontrolled confounding variables --
if smokers tend to be older than the non-smokers and older people are at higher risk of cancer to begin with, then age is a confounder that, if left unadjusted, will lead to an inaccurate estimate of the treatment effect.

The above scenario demonstrates the complexity of causal inference. Analysts often rely on visualizations to inspect and resolve errors in each step of the process.
However, while there are existing visualization libraries used by causal inference experts, they are often not designed specifically for causal inference \cite{Waskom2021, Hunter:2007}, or support only limited analysis tasks \cite{guo2021vaine, causalevaluations, Beaumont_CausalNex_2021}.
This is further complicated by the existence of different causal inference approaches that adopt incompatible assumptions and processes \cite{wang2015visual, wang2017visual, jin2020visual}, resulting in visualization tools that provide only fragmented support for the analysis workflow, and are not designed to work with one another.
Additionally, causal inference often requires multiple repetitions to refine the analysis, but many visualization libraries are not designed for such rapid iteration and do not integrate with the computational environments and statistical packages analysts are familiar with.
Taken together, there are, to the best of our knowledge, no existing visualization toolkits that can be used together to support the various analysis tasks of causal inference practitioners over the entire causal inference workflow.

In this paper, we address these challenges through a design study with causal inference experts.
Working closely with the experts over the course of three months, we first sought to understand the causal inference workflow and analytic tasks through two rounds of formative interviews. We then adopted an iterative design process to develop visualizations to support the tasks, before finally presenting the visualizations back to the experts for feedback and evaluation.
The result of this design study is \textit{Causalvis} (Fig \ref{fig:teaser}), a Python visualization package for causal inference.
Causalvis consists of four visualization modules that support data analysts with tasks such as understanding and communicating causal structure, identifying and statistically controlling for confounding variables, refining cohorts, exploring heterogeneous treatment effects, and tracking analytic provenance.

The contributions of this paper include (i) a characterization of user tasks and design requirements during the causal inference process; (ii) Causalvis\footnote{\url{https://github.com/causalvis/causalvis}}, a Python library that consists of four visualization modules to support experts during causal inference; 
(iii) feedback from experts about the design, functionality, and areas of improvement for each module; and (v) methodological lessons learned from developing and evaluating Python packages with multiple sub-modules in established computational environments.



\section{Related Work}

\subsection{Causal Inference} \label{causal-inference}
{Causality is the overarching paradigm focused on the science of cause and effect.
The methods of uncovering causal structure from data can be further categorized into \textit{causal inference} and \textit{causal discovery}.
Causal discovery (also known as reverse causality \cite{gelman2013ask}) generally aims to infer the causal relations among variables from observational data without specifying treatment or outcome in advance, while causal inference (also known as forward causality) aims to quantify the strength of the causal relationship of a pre-specified treatment on a pre-specified outcome.
For example, researchers would use causal discovery to understand the causal links between variables -- such as smoking behavior, age, biological sex, and lung cancer -- in an observational data set,
and use causal inference to estimate the size of the effect of smoking (treatment) on the risk of lung cancer (outcome) considering influences from covariates.}

{Causal inference, the focus of our work, relies on non-experimental observational data to estimate causal effects.}
Unlike RCTs where participants are randomly assigned to a particular treatment, observational studies do not manipulate an individual's settings or experiences.
Instead, data would be recorded about that individual, such as their demographic information, whether they were exposed to a treatment, and other related attributes.
In a healthcare setting, for instance, researchers can access such observational data in the form of medical databases of patient health and demographic data.
From these data sets, researchers can define the treatment and outcome of interest, and relevant covariates.

The potential outcomes (PO) framework \cite{angrist1996identification}, also called the Neyman-Rubin framework \cite{holland1986statistics} or the Rubin Causal Model \cite{imbens2010rubin}, describes a theoretical framework for causal inference.
Consider again the scenario where we attempt to find the causal effect of smoking on cancer.
To measure the true causal effect of smoking in a population, we need to generate two potential outcomes for every individual -- the outcome had they smoked and the outcome had they not.
We could then quantify the difference between the two outcomes for each individual, and average that difference across the population to obtain the \textbf{average treatment effect} (ATE).
However, in reality, we cannot obtain both potential outcomes for each individual, as each individual either did smoke or did not.
Only one potential outcome is already observed, and the other outcome can never be observed (i.e., counterfactual).
Therefore, causal inference provides the theory and tools for when and how we can estimate causal effects.
Informally, to estimate the ATE, we select a cohort of individuals who share baseline biological/demographic characteristics, 
we divide them into treatment and control groups based on whether they did or did not smoke, perform statistical adjustments to cancel out spurious correlations so the groups are more comparable, and then calculate the difference in average outcomes (i.e., rate of cancer diagnosis) between the two groups.
The PO framework provides the theoretical assumptions required to identify whether this estimated difference from the data at hand is indeed the causal treatment effect.

\ehud{
In addition to the PO framework, there exists a complementary approach to causal inference called the Structural Causal Model (SCM).
SCM relies on the \textit{do}-calculus system pioneered by Judea Pearl \cite{pearl1995causal, pearl2012calculus}, which champions the use of Directed Acyclic Graphs (DAG) for causation.
DAGs consist of nodes and vertices, where nodes correspond to variables in the data and a directed vertex between two nodes indicates a possible causal relationship between the variables \cite{greenland1999causal}.
Generally, DAGs can accomplish two tasks: 1) identifying whether a causal question can be answered from the data, and 2) performing causal effect estimation when used as a computational graph.
Under the SCM framework, practitioners would use DAGs to accomplish both tasks by estimating the effect sizes of all causal relationships (vertices) in a DAG.
In contrast, PO practitioners focus on quantifying the effect of the one vertex connecting the treatment to the outcome, and would use the DAG only for identification (the first task).
This difference in approaches has led the two frameworks to adopt different assumptions and use different sets of estimation tools.
While both frameworks have their strengths and weaknesses, many causal estimation methods that are popular today--such as propensity score methods and matching--were developed under the PO framework \cite{rosenbaum1983central}.
The PO framework has thus seen greater adoption by causal inference practitioners in fields ranging from economics and policy to healthcare and epidemiology.
}


\ehud{
\textbf{Our tool is designed to support the PO framework}, allowing users to estimate the causal effect of a pre-specified treatment on a pre-specified outcome using common statistical tools.
In all subsequent sections, unless otherwise indicated, the term \textit{causal inference} will refer to the PO approach.
}

\subsection{Causal Visualization}

Many visualization techniques have been developed to communicate causal structure and causal relationships, ranging from static directed acyclic graphs~\cite{pearl2000models, rohrer2018thinking} to animated visualizations~\cite{elmqvist2003growing, elmqvist2003causality, kadaba2007visualizing}.
More recently, studies have looked at combining textual narratives with causal graphs to help users understand temporal events~\cite{choudhry2020once}, as well as how visualizations might be leveraged for causal support~\cite{kale2021causal}, interpreting counterfactuals~\cite{kaul2021improving}, identifying mediator variables~\cite{yen2019exploratory}, and conveying causality in biological pathways~\cite{dang2015reactionflow}.
Studies have also explored how visualizations might erroneously convey an illusion of causality in data~\cite{xiong2019illusion}.

An important prerequisite for causal inference is a directed acyclic graph (DAG) depicting the possible causal relationships between different variables in the data.
Traditionally, these causal relationships would be manually specified by a domain expert, such as a doctor or physician, aided by tools like Dagitty \cite{textor2016robust} that enable interactive specification of DAGs.
More recently, automatic methods for \textit{causal discovery} were developed to recover causal structures and learn causal relationships between variables from observational data \cite{glymour2019review}.
These tools can be fully automated, like Causalnex \cite{Beaumont_CausalNex_2021}, 
or interactive, like SeqCausal \cite{jin2020visual} and DOMINO \cite{wang2022domino}, enabling humans to take part in discovering the underlying causal structure from sequential data.
However, in both cases, causal discovery tools do not quantify the effect of a treatment on an outcome variable, and require subsequent use of \textit{causal inference} methods.

\ehud{
There exists a broad range of visualization tools that have been developed for causal inference, such as visualizing and refining causal structures \cite{wang2015visual}, performing analyses \cite{wang2017visual}, explaining the AI models used \cite{hoque2021outcome}, and debiasing AI algorithms \cite{ghai2022d}.
While these visualization tools support a variety of user tasks,
they are mostly grounded in the SCM framework and are not compatible with in the more popular PO framework.
For example, while both frameworks might use DAGs to visualize causal relationships, the analytic goals of the visualization would differ.
Previous approaches, such as the Visual Causality Analyst \cite{wang2015visual} and the Causal Structure Investigator \cite{wang2017visual}, use multiple regression models to estimate the effect size of all causal relationships in the DAG (see \ref{causal-inference}).
It thus makes sense for these tools to encode effect sizes using vertices between nodes.
In contrast, the PO framework focuses on estimating the effect size of a single predefined treatment on a predefined outcome (i.e., a single vertex).
This smaller scope allows analysts to use more diverse estimation methods that are commonly available in statistical tools, but also requires analysts to adjust for variables that may bias the estimation.
A DAG built for the PO framework would thus need to highlight how variables in the graph relate to the treatment-outcome relationship (e.g., by introducing confounders), a task that is not supported by the previous SCM-based visualization tools.
}

\add{This example illustrates only one of many} distinctions between the two approaches.
\add{These distinctions} mean that visualizations designed for the PO framework must support entirely different user tasks and analysis processes.
To this end, we identified only three visualization packages developed for the PO framework -- VAINE~\cite{guo2021vaine}, causallib \cite{causalevaluations}, and Cobalt~\cite{greifer2020covariate}.
VAINE and causallib are both designed for use with the Jupyter notebook environment. 
VAINE is an interactive visual analytics tool that helps users identify clusters in the data set and estimate the average treatment effects across clusters, while causallib uses static visualizations to help users evaluate their causal inference models.
In contrast, Cobalt is a visualization package designed for R, and helps analysts validate that their selected samples are suitable for causal inference.
Of these, causallib and Cobalt only provide static visualizations, which makes rapid iteration and interactive analysis time-consuming.
Furthermore, each package only addresses limited tasks in the causal inference process, and some tasks, such as identifying variable types, are not supported in any tool.
\bc{
To address this gap, we conducted a design study with experts to understand users' workflows in causal inference and to build and evaluate an interactive visualization system to support their analysis.
}

\subsection{Visualizations in Computational Environments}
Computational notebooks (e.g., JupyterLab, Google Colab, and Kaggle Notebooks) are programming environments in which users can interweave segments of code and output within the same interface.
These notebooks have been widely adopted for their ability to support exploratory data analysis \cite{subramanian2020casual, kery2018story}, collaboration \cite{zhang2020data}, rapid iteration and workflow documentation \cite{wang2021makes}.
Recent works have advocated for the development of interactive visualizations in such computational environments in order to support reproducibility, streamline analysis and increase adoption of visualization systems \cite{ono2021interactive, wang2022nova}.
To support these goals, tools have been developed that help users embed interactive visualizations \cite{wang2022nova}, create dashboards \cite{wang2022stickyland, wu2020b2}, and reuse workflows \cite{gadhave2021reusing} in JupyterLab.
Studies have also developed tools to condense notebooks for better collaboration \cite{rule2018aiding} and communication through interactive data comics \cite{kang2021toonnote} and presentation slides \cite{zheng2022telling}.
In addition to these general purpose tools, there has also been a trend towards fluid, interactive widgets embedded within notebook environments \cite{kery2020future}.
Within data visualization, a range of interactive notebook widgets have been developed for use in domains such as biology \cite{fernandez2017clustergrammer}, machine learning \cite{palmeiro2022data, bauerle2022symphony}, data comparison \cite{wang2022diff}, and programming education \cite{horn2022tunepad}.
Specific to causal inference, both VAINE \cite{guo2021vaine} and causallib \cite{causalevaluations} are designed to be used in Jupyter Notebook.
Inspired by these tools, we implemented Causalvis as a Python package for computational notebooks so that users can easily collaborate with experts and iterate through their causal inference workflow rapidly.

\section{Design Study} \label{formative}

To understand the process of causal inference, users' tasks, and their analytic goals, we conducted a design study \cite{sedlmair2012design} with eight causal inference experts.
Through two rounds of interviews, we first performed a formative user task analysis to derive the typical causal inference workflow and relevant analytic tasks.
We then validated our findings in a second round of interviews, where the same experts helped refine and elaborate on the workflow and ideate through low-fidelity wireframes of visualization designs that can support their work.
In the following sections, we provide a brief background on the causal inference framework, and describe the participant recruitment and design study process.
Then, we present the causal inference workflow and user analytic goals that were derived from the interviews.

\subsection{Participants}
The target users of our system are causal inference experts who are interested in using causal inference to estimate the effect of a treatment on an outcome within a particular usage domain.
They are familiar with the causal inference process and are experienced using it in prior or current projects.

We recruited experts through a snowballing method. We first posted a message on a Slack channel, and reached out to project managers using our enterprise network.
Through these connections, we branched out and recruited other participants with relevant expertise in causal inference.
A total of eight experts (E1-8) in diverse domains were recruited.
Participants were asked to self-report on a scale of 1 (no experience) to 5 (I consider myself an expert) their proficiency in Python ($\mu$=3.86, $\sigma$=0.690), JupyterLab ($\mu$=3.29, $\sigma$=1.38), causal inference ($\mu$=4.14, $\sigma$=1.07), creating/using DAGs ($\mu$=3.71, $\sigma$=0.76), and data visualization ($\mu$=3.86, $\sigma$=0.690).
One participant declined to provide the above information.
Three of the participants are consultants who serve as contacts for clients interested in causal inference. Their clients have domain expertise and relevant data sets, while their role as consultants is to perform causal analysis and communicate results back to the clients.
Two of the participants are data scientists and researchers involved in the development of libraries for causal inference. They are also domain experts who have used causal inference in healthcare research, and are experienced in conducting and reporting research results.
The other three participants are graduate students who work on causal inference, including theoretical (non-domain specific) simulations as well as healthcare (domain specific) related applications.

\subsection{Study Procedure}
\label{sec:study_procedure}
To understand the process of causal inference and tools used by experts, we first conducted formative interviews with causal inference experts. 
The interviews were semi-structured, and
where applicable, we asked specific follow-up questions based on user responses and their area of expertise. 
Each session lasted for no more than an hour, with 1-3 causal inference experts on each call.
From these interviews, we gained an initial understanding of the sequence of tasks typically performed during the causal inference process and the challenges faced.
We also obtained an overview of the data domains the experts worked in, as well as the current ecosystem of tools and libraries that support their work.

Next, we created an initial three-step workflow summarizing the causal inference process.
We also made low-fidelity wireframes of possible visualization tools that might be used to support each step of the workflow.
These wireframes were simple prototypes, sketches, or screenshots of visualizations that have been published in causal inference research papers.
For each visualization design, we provided multiple alternatives and annotated the images to indicate how interactions would work.
We also combined them with screenshots of pseudo-code written in JupyterLab to better reflect how the designs would integrate into the computational environment as a visualization module.

Finally, in a second round of interviews, we presented the workflow and wireframes back to our experts for validation and feedback.
During these interviews, specific comments were surfaced to improve our understanding of the workflow.
Participants also pointed out particular features and changes that can be implemented in the visualization wireframes.
Building on the feedback, we then refined the three-step workflow to better reflect the causal inference process (see Section \ref{workflow}).
We also came up with a set of design goals for Causalvis that capture the needs and requirements of our users (see Section \ref{DG}).
The workflow and design goals are presented in the following sections.


\subsection{The Causal Inference Workflow} \label{workflow}

The causal inference workflow presented here summarizes the process shared by all the causal inference experts we interviewed (Fig. \ref{fig:workflow_diagram}).
The causal inference workflow begins with an observational data set and can be summarized into 3 main steps: 1) Causal Structure Modeling, 2) Cohort Construction/Refinement, and 3) Treatment Effect Exploration. 
The three steps are described in detail below.
We present experts' remarks in quotes and italics where appropriate.

\subsubsection{Causal Structure Modeling} \label{causal_structure_modeling}
The first step of causal inference is typically causal structure modeling.
The goal of this step is for causal inference analysts to accurately model the causal relationships between variables in the data set and identify the variables that must be adjusted for during the analysis process.

\begin{figure}[h]
  \centering
  \includegraphics[width=\linewidth]{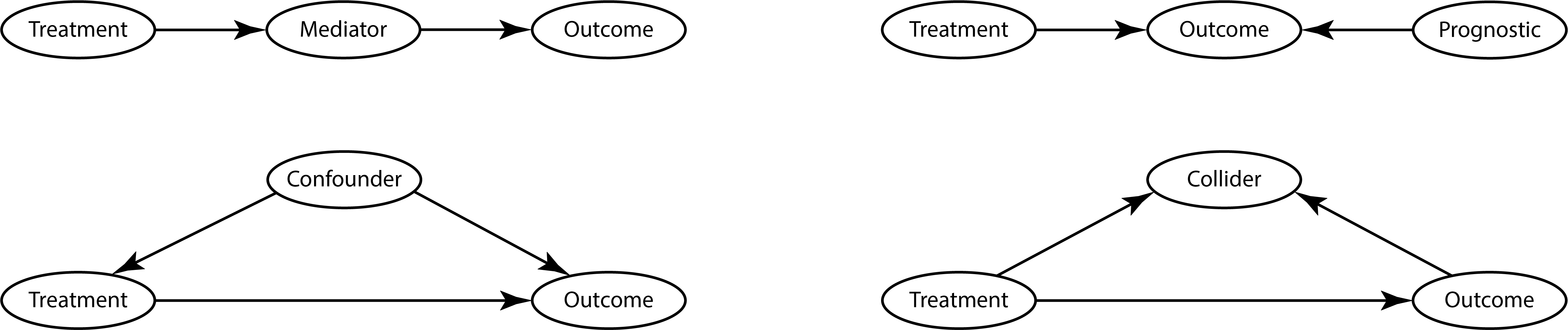}
  \caption{Common variable types and their relationship to the treatment and outcome variables.}
  \label{fig:DAG-types}
  \Description{Four directed acyclic graph diagrams laid out 2 by 2. From top-left going in a clockwise direction, they depict the variable relationships: mediator, prognostic, collider and confounder.}
\end{figure}

In this step, causal inference analysts often \textit{``start with a causal graph''} (E3).
Causal graphs are \textbf{directed acyclic graphs} (DAGs) where nodes are variables, and a directed vertex from node M to N indicates that M is a likely cause of N.
From this graph, analysts would identify the variables to adjust in order to satisfy the assumption that there are no unmeasured confounding factors.
Confounders are variables that affect both the treatment and the outcome, and if left unadjusted, they can introduce bias to the treatment effect estimation. 
An estimated treatment effect is valid only if all confounders are identified and adjusted for.
In addition to confounders, there are a few other variable types that must be identified: \textbf{mediators}, \textbf{colliders}, and \textbf{prognostic factors}, to name the most common ones \cite{hernan_robins_2020} (see Fig. \ref{fig:DAG-types} for graphical representations of these variable types).
Not all variable types should be adjusted for, and causal inference analysts \textit{``care a lot about the variables''} (E2) that are used in subsequent analyses.
Adjusting for colliders and mediators (aka post-treatment variables), for example, might bias the estimation instead.

Accurately modeling the causal structure and identifying different variable types is thus a crucial outcome of this step.
In our formative interviews, causal inference experts mentioned that they typically collaborated with domain experts who have \textit{``some intuition about what is relevant''} (E3) and what the causal relationships are between variables.
However, domain experts do not always have the causal inference expertise needed to specify such a graph from scratch.
They may not have the knowledge \textit{``in terms of what's a confounder or what's a mediator''} (E1), and asking them to express their intuition as a DAG can be difficult and time-consuming.
Causal inference analysts may thus rely on their own knowledge or use a causal discovery package, such as Causalnex \cite{Beaumont_CausalNex_2021}, to first create a ``draft'' of the DAG, which is then presented to collaborators to iteratively validate links and refine the graph.
However, the experts we spoke to wanted a system to better support this process, facilitate \textit{``interactively building a causal graph''} (E2) and serve as \textit{``a basis for conversation''} (E5) with their domain expert collaborators.
Only after they have validated that the causal structure is represented accurately do analysts use the DAG to identify variables that need to be adjusted for in subsequent steps of causal inference.

\subsubsection{Cohort Construction/Refinement} \label{cohort_construction}
The next step of causal inference is to select a cohort of treatment and control samples from the observational data set to estimate the effect of the treatment on the outcome variable.
For a cohort to yield valid causal effect estimations, it should satisfy the \textbf{positivity} assumption. 
Informally, this assumption validates that the two treatment and control groups in the cohort are not too distinct from one another.
More formally, it states that each sample must have a positive probability to be in either treatment condition, meaning that the covariate distributions of the two treatment and control groups should overlap.
This can be either forced or tested, depending on the model the analysts apply to estimate the treatment effect.

There are two main modeling approaches used for this step: matching and inverse propensity weighting.
\textbf{Matching} \cite{rubin1973matching} is an intuitive method to create comparable treatment groups,
matching each individual in the treatment group with an individual from the control group who has similar covariate values.
This method, by construction, usually forces the groups to be compatible, but can discard many samples in the process.
Another common method that uses the entire data set to estimate the ATE is \textbf{inverse propensity weighting} (IPW) \cite{rosenbaum1983central}.
The \textbf{propensity score} is the probability of a unit to be assigned to the treatment group.
The inverse of the propensity score can then be used to create a weighted pseudo-population in which the distribution of characteristics is similar between groups.
Since we use the entire cohort, and the treatment assignment is not randomized, the positivity assumption is not guaranteed to hold. 

In our formative interviews, experts mentioned that checking for positivity violations was often the first thing they did regardless of method or data domain because \textit{``if you have a positivity issue, you cannot work on this cohort''} (E5).
Experts typically performed this by comparing the propensity score distributions between groups \cite{causalevaluations} since in theory, the propensity score is a sufficient summary of the covariate space \cite{rosenbaum1983central}.
If the propensity score distributions of the treatment and control groups did not overlap, analysts would identify and exclude the imbalanced samples from the cohort, iteratively refining the cohort until the groups were well-balanced.

\subsubsection{Treatment Effect Exploration} \label{treatment_effect_exploration}
In the final step of causal inference, analysts want to explore and identify heterogeneous treatment effects in the cohort.
Some causal effect estimation methods, such as matching, allow individual treatment effects to be calculated for each sample in the data set, which in turn lets analysts \textit{``aggregate over the treatment groups you care about, for example males and females, old and young''} (E2).
Analysts can then explore how the average treatment effects for each subgroup differs from the others and \textit{``find these highly differentiated subgroups''}.
When the conditional treatment effect varies between different levels of the data, this is an indication of a \textbf{heterogeneous treatment effect}.
Identifying heterogeneous treatment effects is an important task for many of the causal inference experts we spoke to because estimating the ATE for the entire cohort did not always provide enough insight into the causal effect.
Instead, comparing distinct subgroups can help them better understand the data, such as identifying the populations for whom a treatment would be most effective or examining outcomes for \textit{``populations of different characteristics''} (E5).

\begin{figure*}[h]
  \centering
  \includegraphics[width=\textwidth]{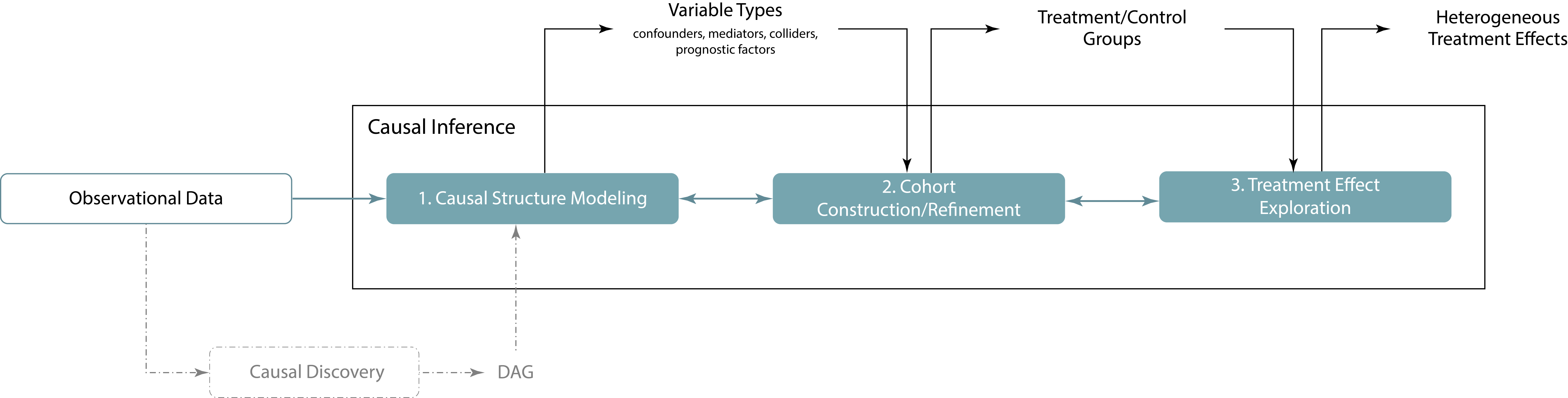}
  \caption{The causal inference workflow begins with some observational data and consists of 3 main steps: 1) Causal Structure Modeling, 2) Cohort Construction/Refinement, and 3) Treatment Effect Exploration. Double-ended arrows indicate iterative steps where analysts return to an earlier process to refine their analysis. The inputs and outputs are indicated above the corresponding steps. For researchers, the goal of each step is to obtain the outputs, which may then be passed on as inputs in subsequent analyses. We also include an optional causal discovery step that may sometimes be used to automatically generate initial DAGs from the data set that are later refined by domain experts during Causal Structure Modeling.}
  \label{fig:workflow_diagram}
  \Description{A five step model. The three steps of the causal inference workflow are colored and enclosed in a box. Two-way arrows connect the steps of the causal inference workflow. The observational data and causal discovery steps are outside the box and connected by uni-directional arrows to the causal inference steps. Above each causal inference step, the input/output of the step are listed and connected by arrows.}
\end{figure*}

\subsection{User Analytic Tasks} \label{DG}

Based on the two rounds of interviews with our causal inference experts and the user workflow described earlier, we found that participants often worked in JupyterLab, and would like to stay within the computational environment \textit{``for the convenience''} (E1).
Additionally, we also derived eight analytic tasks that should be supported by our visualization package.
For each task, we indicate if it is only relevant to one of the steps in the workflow described above.
These tasks were then used to guide the design of the Causalvis visualization modules.



\begin{enumerate}
    \item[\textbf{T1}] \textbf{Collaboratively creating and communicating causal structure.} [Causal Structure Modeling] Many experts we interviewed mentioned that causal inference is often a collaborative effort that requires the input of domain experts who are familiar with the data, but may lack expertise with causality. Causalvis should thus make it easy to communicate the causal structure of a data set. It should also be easy to create and modify the causal structure without technical expertise or in-depth knowledge about causality. \label{collaboration}
    \item[\textbf{T2}] \textbf{Maintaining the independence of causal structure from specific data sets.} [Causal Structure Modeling] Discussion between data scientists and domain experts over causal structure can sometimes begin before data sets are available as it might guide what variables are even needed in the first place. They may also want to include additional variables that are known to be relevant. Causalvis should thus \textit{``not be constrained to the data set itself''} (E1) and support the modeling of causal structures independently of any data\label{independence} set. 

    \item[\textbf{T3}] \textbf{Identifying different types of variables.} [Causal Structure Modeling] Given the importance of selecting the right control variables during causal inference, variable types, such as confounders, mediators, colliders and prognostic factors, should be emphasized as \textit{``something to be mindful of''} (E3). Causalvis should help users quickly identify the different variable types and keep track of them. \label{different_variables}

    \item[\textbf{T4]}] \textbf{Checking covariate balance and positivity violations.} [Cohort Construction/Refinement] Ensuring that treatment and control groups are comparable during cohort refinement is a crucial task for obtaining unbiased estimates of the treatment effect. Causalvis should help users identify when covariate balance and the positivity assumption are violated. 
    Furthermore, Causalvis should allow users to select samples that should be excluded in order to satisfy the positivity assumption. \label{positivity}

    \item[\textbf{T5}] \textbf{Estimating treatment effects conditioned on a variable.} [Treatment Effect Exploration] It is often insufficient to estimate a treatment effect for an entire population. Data analysts want to condition on certain variables and gain insight into how the treatment effect differs across subgroups. Causalvis should support the exploration and identification of heterogenous treatment effects in a cohort. \label{heterogenous}
    \item[\textbf{T6}] \textbf{Supporting a flexible and iterative workflow.} Common across all our experts is the highly flexible and iterative nature of their causal inference process. Experts often described the analysis steps as being ``iterative''. Different experts working in different domains may also use different causal inference methods, skip steps in the workflow, or prioritize one stage while using heuristics in the others. The three steps identified in \ref{workflow} are thus neither prescriptive nor unidirectional. Causalvis should allow users to iterate through each step and return to an earlier step to refine their process when needed. \label{iteration}
    \item[\textbf{T7}] \textbf{Tracking analytic provenance.} Since there is often no established ground truth when conducting causal inference on observational data, analysts often iterate through multiple hypotheses, cohorts, and estimates of the treatment effect value. They were thus \textit{``in favor of version control''} (E5). Causalvis should provide a method of tracking and comparing outcomes across different analytic decisions. \label{provenance}
    \item[\textbf{T8}] \textbf{Integrating with existing causal analysis packages.} From our formative user interviews, we found that there was no one unified method of performing causal inference, and experts expect Causalvis to integrate seamlessly with existing data formats (such as networkx\footnote{https://networkx.org/} graphs) and libraries (such as CausalNex \cite{Beaumont_CausalNex_2021} and causallib \cite{causalevaluations}) that they already use for causal inference. \label{integration}
\end{enumerate}

\section{Causalvis}

Based on the user workflow (see \ref{workflow}) and analytic tasks (see \ref{DG}) identified from our design study, we developed the Causalvis visualization package to support analysts through the three steps of the causal inference process.
The package is developed for use in the JupyterLab computational environment since all experts in our formative study mentioned that they typically worked with Python packages and JupyterLab notebooks.
We took an iterative design approach to our development process. In addition to the two rounds of formative interviews described, we also presented videos of the modules to the experts during implementation, informally collecting intermediate feedback to refine our designs.

The Causalvis package consists of four visualization modules, where the first three modules (DAG, CohortEvaluator, and TreatmentEffectExplorer) correspond to the three steps identified in the causal inference workflow and the fourth module (VersionHistory) is designed to track analytic provenance.
They are designed to work independently of one another \textbf{(T6)}.
A key choice made in our design process was to focus on the visualizations needed for each step (see Fig. \ref{fig:workflow_diagram}), while leaving the data processing and statistical adjustment techniques at the discretion of the user.
We thus emphasized the ability for each module to integrate seamlessly with existing libraries and data structures instead.

The Causalvis front-end is implemented in Javascript, using the React framework\footnote{https://reactjs.org/}.
All visualizations are implemented using D3.js\footnote{https://d3js.org/}.
Each module is integrated into the JupyterLab\footnote{https://jupyter.org/} computational environment using Python and the IPywidget framework\footnote{https://ipywidgets.readthedocs.io/en/stable/}.
In the following sections, we describe the modules in detail.
Where relevant, we indicate the \function{functions} available in each module, the \argument{arguments} accepted, and the \variable{variables} that can be accessed from the module class.
For each module, we also discuss visualization packages currently available to analysts. We describe how Causalvis innovates and extends these existing solutions, with emphasis on the additional tasks supported by our modules, informed by formative interviews with causal inference experts.

\subsection{Usage Scenario} \label{usage-scenario}

When describing the visualization modules in the following sections, we will simultaneously walk through a usage scenario \cite{isenberg2013systematic} to demonstrate and contextualize how Causalvis might be used in a realistic causal inference task.
We use the UCI Student Performance data set \cite{cortez2008using} in this scenario\footnote{We used the data set for Mathematics.}.
This data set records student math grades and related data at two Portuguese schools throughout the year. 
There are 30 attributes that track student demographic, social and academic information (such as \textit{age}, \textit{address}, and \textit{absences}), and 3 attributes that track student grade throughout the year \textit{(G1, G2, G3)}.
We drop 7 sensitive or open-ended attributes from the data set \textit{(school, sex, age, Mjob, Fjob, reason, guardian)}, leaving a total of 26 attributes and 395 samples (rows).

This data set has previously been used to build machine learning models predicting student academic performance \cite{cortez2008using} and demonstrate the use of causal discovery packages such as Causalnex \cite{Beaumont_CausalNex_2021}.
We select it for our Causalvis usage scenario because the data domain is intuitive, and the existence of prior use cases serves as a baseline for causal inference analysis.
The treatment attribute of interest is \textit{absences}, found to be a predictive factor in machine learning models of student performance \cite{cortez2008using}.
However, the size of its causal effect has not been estimated in prior studies.
In this usage scenario, we convert \textit{absences} to a binary variable based on the data set median, such that students with $\geq median$ absences is encoded as having frequent absences (\textit{absences} = 1), while students with $< median$ absences have few or no absences (\textit{absences} = 0).
The outcome attribute of interest is \textit{G1}, the grade obtained in the first exam of the year.

The scenario will follow the three-step workflow introduced in \ref{workflow}.
Throughout the example, we may use other packages, such as causallib, to perform the statistical calculations necessary for causal inference.
These external packages are not requirements or dependencies of Causalvis.
Analysts may use other statistical packages they are familiar with so long as the data is passed to Causalvis in the expected data format.

\subsection{DAG}

\begin{figure*}[h]
  \centering
  \includegraphics[width=\textwidth]{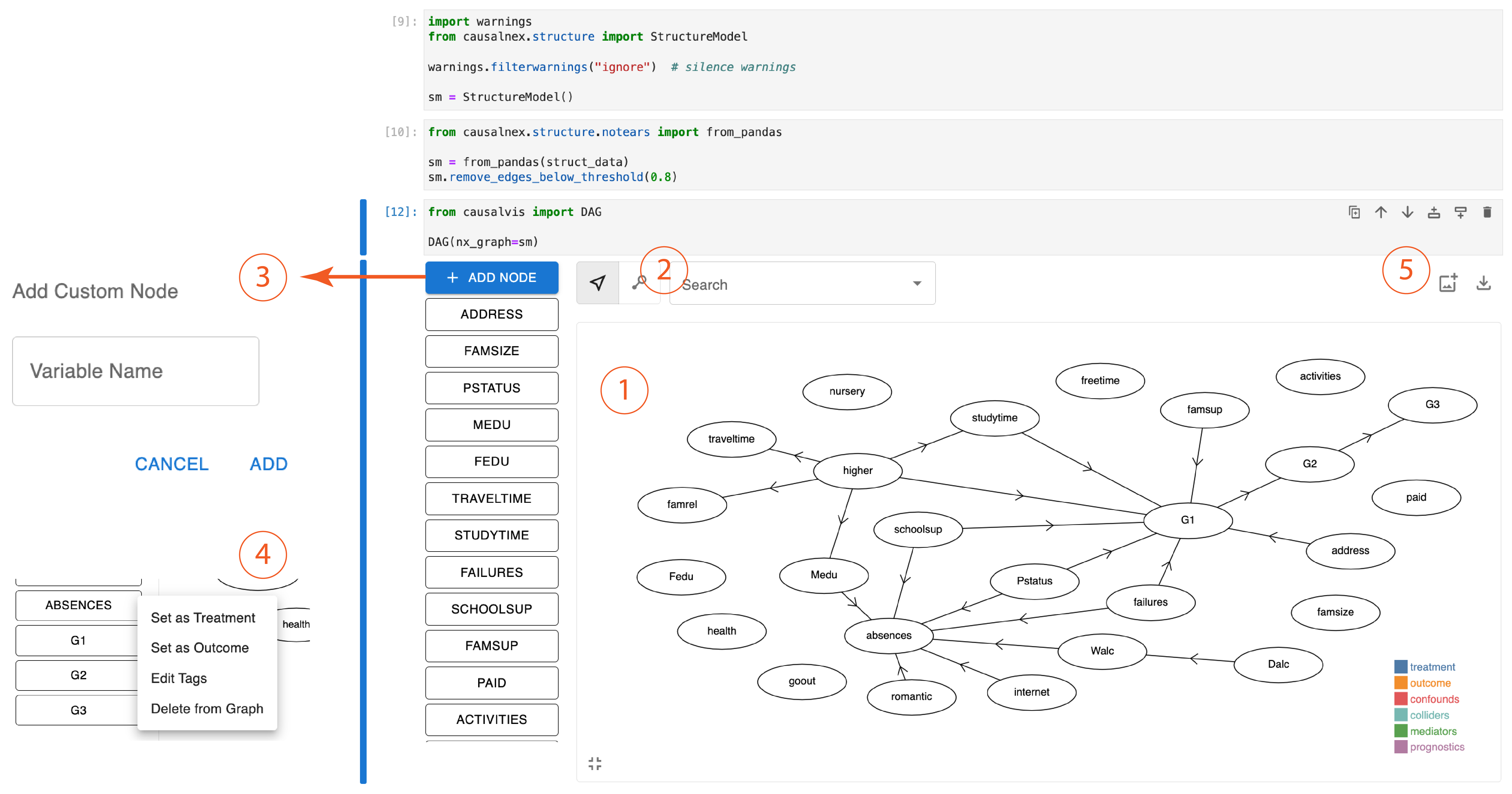}
  \caption{The DAG module initialized with a networkx graph. 1) The graph is visualized on load. 2) Toggle buttons can be used to switch between layout editing and link editing. 3) The \textit{Add Node} button can be used to add custom nodes to the DAG. 4) The context menu of each variable can be used to set treatments and outcomes, edit tags, and delete variables from the visualization. 5) The DAG can be downloaded and saved as an image or \textit{.json} file.}
  \label{fig:DAG_example}
  \Description{A screenshot of the DAG module. To the left of the screenshot are two callout boxes showing the add custom node dialog and the context menu for each variable. Parts of the screenshot are labeled from 1-5.}
\end{figure*}

The DAG module (Fig. \ref{fig:DAG_example}) is designed to help users quickly and effectively model different causal structures using directed acyclic graph (DAG) visualizations.
In the Causal Structure Modeling step (see Section \ref{causal_structure_modeling}), users want to understand the causal relationships and identify the variables that need to be adjusted for.
They typically collaborate with domain experts to iteratively refine the DAG and construct different hypotheses.
While some existing packages, such as Causalnex (see \ref{app_causalnex}), allow users to model causal structures by specifying nodes and links between nodes, this can only be done manually by writing code, and the resulting DAG visualizations cannot be interactively edited through direct manipulation.
This makes the process of creating and refining DAGs too time-consuming, particularly when collaborating through infrequent meetings with domain experts who may not have the time or technical expertise to refine DAGs programmatically.
The DAGitty \cite{textor2016robust} (see \ref{app_dagitty}) application is a tool that supports the interactive editing of DAGs. However, the application expects users to know causal structure terminology (e.g. confounders, conditional independence), which may be unfamiliar to domain experts who are often not causal inference experts.
Our Causalvis DAG module extends these existing tools by supporting interactive DAG modeling within the JupyterLab computational environment itself.
By allowing users to create and refine DAGs directly on a visual interface, we facilitate the interactive and collaborative causal structure modeling process needed by causal inference analysts (see \ref{causal_structure_modeling}, \textbf{T1}).
Additionally, we implement automatic variable type identification for subsequent analysis, and image download features that support easy sharing and communication with subject matter experts \textbf{(T1)}.

\textbf{Initializing the DAG Module.} The DAG module can be initialized in a variety of ways, the simplest of which is using \function{DAG()} to create an empty canvas without first obtaining an existing data set \textbf{(T2)}.
Additionally, the module also accepts the following arguments: \argument{attributes}, \argument{graph}, \argument{data}, and \argument{nx\_graph}, allowing users to flexibly create and edit DAGs based on the data input available.
These different input formats are included to support collaboration between users, allowing analysts to quickly load causal graphs that have been created beforehand \textbf{(T1)}.
Some input formats also help integrate the DAG module with existing Python packages \textbf{(T8)}.
For instance, the Causalnex package outputs causal structures in the networkx data format, which can be directly passed into the module using the \argument{nx\_graph} argument.

\textbf{Creating and Editing DAGs.} When creating or editing a DAG, users can add nodes by clicking on the variable name from the list on the left.
If the module is initialized with no variables, or if users wish to capture the relationship of additional factors, new nodes can be added to the canvas using the \textit{Add Node} button (Fig. \ref{fig:DAG_example} \textcircled{3}). This allows users to quickly and flexibly capture domain knowledge about causal relationships without being restricted by any existing data set \textbf{(T2)}.
Users can also toggle between the \textit{move} and \textit{edit links} buttons to either reposition the layout of the nodes or edit the links in the DAG (Fig. \ref{fig:DAG_example} \textcircled{2}). By supporting direct interaction with the DAG, users can iterate rapidly through different hypotheses \textbf{(T6)}, and subject matter experts can collaborate on modeling the causal structure even without technical programming know-how \textbf{(T1)}.



\textbf{Identifying Variable Types.} Each variable in the list has a context menu that has the following options: \textit{Set as Treatment}, \textit{Set as Outcome}, \textit{Edit Tags}, and \textit{Delete from Graph}.
The \textit{Set as Treatment} and \textit{Set as Outcome} options will designate a particular variable as either the treatment or outcome of interest and will change the color of the corresponding node in the DAG (Fig. \ref{fig:DAG_example} \textcircled{4}). There can only be one treatment and one outcome in each DAG, and a single variable cannot be both.
Furthermore, when both treatment and outcome have been selected, all other nodes will be automatically colored to highlight the different variable types: \textit{confounders}, \textit{colliders}, \textit{mediators}, and \textit{prognostics}.
Variable types are identified by recursively traversing the target-source relationships in the node-link structure (Fig. \ref{fig:DAG-types}), and are dynamically updated whenever the user edits the DAG or the treatment and outcome variables.
These highlights can help users identify the variables that must be adjusted for in subsequent analyses \textbf{(T3)}.

\textbf{Saving and Sharing DAGs.} Once a DAG has been created, users can share the causal model by downloading the DAG as a \textit{.png} image using the \textit{download image} button \textbf{(T1)} (Fig. \ref{fig:DAG_example} \textcircled{5}). Alternatively, the node-link structure can also be shared as a \textit{.json} file using the \textit{download json} button.
This file can be customized to include information about the different variable types identified.

\textbf{Accessing DAGs and Variable Types in Python.} Data analysts who wish to use the outputs of the DAG module in subsequent analyses can also quickly access the relevant data variables in the Jupyter notebook without downloading the \textit{.json} file \textbf{(T8)}.
The causal structure created can be obtained using \variable{.DAG}, and the different variable types can be accessed similarly with \variable{.confounds}, \variable{.colliders}, \variable{.mediators}, and \variable{.prognostics}.
These variable types can then be used in subsequent analysis to determine the statistical adjustments that need to be made \textbf{(T3)}.


\begin{figure}[h]
  \centering
  \includegraphics[width=\linewidth]{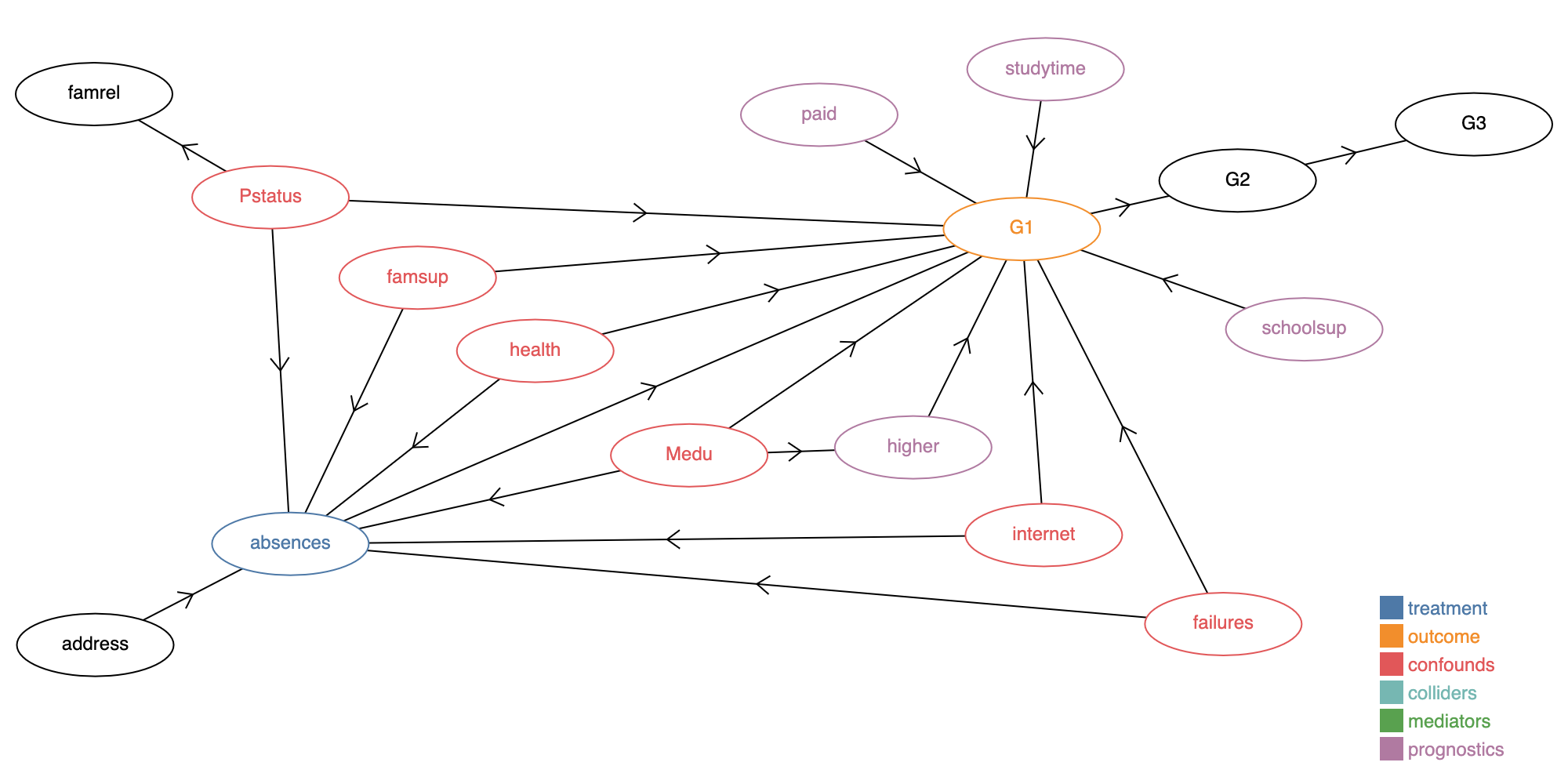}
  \caption{The DAG of the causal relationships in the student learning data set, refined from Figure \ref{fig:DAG_example}. This image was downloaded directly from the Causalvis DAG module using the \textit{download image} function and presented without further processing.}
  \label{fig:student_DAG_raw}
  \Description{The DAG of the causal relationships in the student learning data set. Some nodes are highlighted in color. A legend in the bottom right depicts the mapping from color to variable type. The variable types are: treatment, outcome, confounds, colliders, mediators and prognostics.}
\end{figure}

\subsubsection{Usage Scenario}
We start by constructing the DAG model of the Student Performance data set to visualize the expected causal relationships between different attributes.
To do so, we can load the data set into a \verb|DataFrame| using the Python pandas package, and pass this to the DAG module to manually create a DAG from scratch.
More effectively, however, we can also leverage the Causalnex package to automatically create an initial `discovery' of what the causal structure should be.
We then load the Causalnex graph into the DAG module (Fig. \ref{fig:DAG_example}) to delete spurious nodes and links, add connections based on domain knowledge, and identify variable relationships.
After iteratively editing the graph, we obtain a revised version of the DAG that represents our hypothesis of the causal structures that affect student absence and exam grades.
We then set \textit{absences} as the treatment variable, and \textit{G1} as the outcome variable.
The DAG automatically updates to highlight the other relevant variable types (Fig. \ref{fig:student_DAG_raw}).
From this, we can identify that there are six confounding variables \textit{(Pstatus, famsup, health, Medu, internet, and failures)} and four prognostic variables \textit{(paid, studytime, schoolsup, higher)} that should be adjusted for in subsequent steps of the causal inference analysis.


\subsection{Cohort Evaluator}

\begin{figure*}
  \centering
  \includegraphics[width=\textwidth]{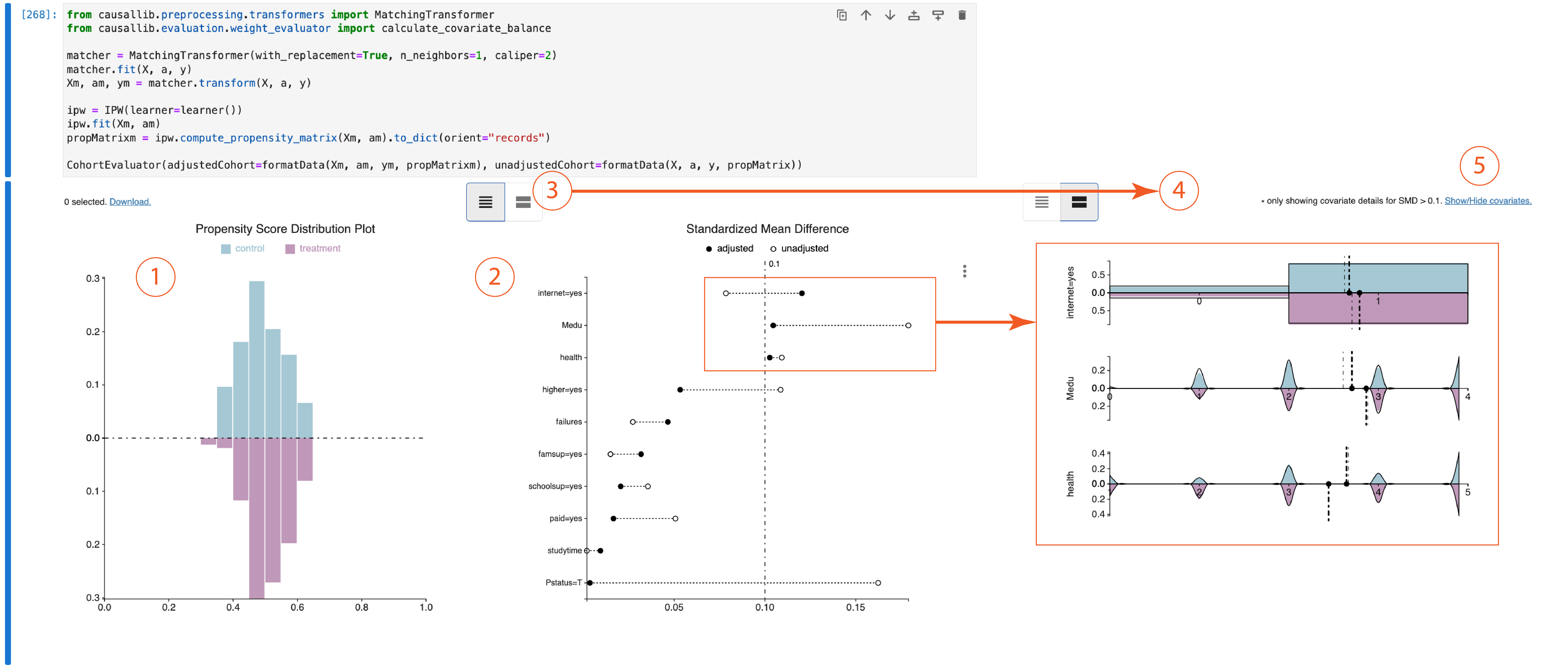}
  \caption{The CohortEvaluator module initialized with both adjusted and unadjusted cohort arguments. 1) The propensity score distribution of the treatment and control groups. 2) The absolute Standardized Mean Difference (aSMD) plot visualizing the aSMD of each covariate. 3) Toggle buttons can be used to view detailed distributions of each covariate. 4) The detailed distributions view automatically shows the distributions of covariates with an adjusted aSMD above 0.1. 5) The \textit{Show/Hide covariates} button can be used to customize which covariate distributions are shown, including already well-balanced covariates.}
  \label{fig:ceval}
  \Description{A screenshot of the CohortEvaluator module. To the right of the screenshot is a callout box showing the detailed covariates view. Parts of the screenshot are labeled from 1-5.}
\end{figure*}

The CohortEvaluator module (Fig. \ref{fig:ceval}) is designed to help users validate that their selected cohorts satisfy positivity assumptions, and to refine the cohorts when necessary (see \ref{causal_structure_modeling}).
When conducting causal inference analysis, the covariate distributions of the treatment and control groups should be as similar as possible to reduce the effect of biasing covariates.
The propensity score plot and absolute Standardized Mean Difference plot (aSMD plot, also called Love plot) included in this module have been widely used in causal inference \citep{granger2020review}, and are also part of causal inference toolkits like causallib \cite{causalevaluations} in Python and Cobalt \cite{greifer2020covariate} in R.
In causallib (see \ref{app_causallib}), these visualizations are only supported when the IPW method is used, which can exclude the use of other approaches, such as matching (see \ref{cohort_construction}).
Cobalt (see \ref{app_cobalt}) supports both methods, but has no Python or Jupyter notebook equivalent.
Furthermore, although Cobalt offers customization of the propensity score and covariate distribution visualizations, only static visualizations are available, and each plot must be generated through separate function calls.
This process can thus be considerably time-consuming, particularly when many iterations of cohort refinement are required.
The Causalvis CohortEvaluator module addresses these gaps with interactive propensity score and aSMD plots that allow imbalanced samples to be selected and excluded from subsequent iterations of cohort refinement.
Detailed distributions of imbalanced covariates are also visualized automatically.
Finally, the entire module is implemented for the Jupyter notebook environment.

\textbf{Evaluating Propensity Balance.} The propensity score plot visualizes the propensity score distribution of the treatment and control groups (Fig. \ref{fig:ceval} \textcircled{1}). In a cohort that satisfies the positivity assumption, these two distributions should overlap. 
Lack of overlap suggests that some individuals in the population are not comparable since they don't have a valid counterpart in the other exposure group \textbf{(T4)}.

\textbf{Identifying Positivity-Violating Samples.} If there is a lack of overlap between treatment and control group distributions in the propensity score plot, it is often necessary to identify these individuals and exclude them from the cohort.
This can be done by brushing over the propensity score plot to select the violating subpopulation.
The selected samples can then be downloaded using the \textit{Download} button and excluded from the cohort such that the positivity assumption holds for all subsequent analyses \textbf{(T4)}.
In addition to downloading the selection, samples can also be accessed in the Python notebook using \variable{.selection}.
The inverse can similarly be accessed using \variable{.iselection}.

\textbf{Evaluating Covariate Balance.} The aSMD plot is another visualization often used by causal inference experts to validate that treatment and control groups are well-balanced (Fig. \ref{fig:ceval} \textcircled{2}). In a well-balanced cohort, the standardized mean difference between the treatment and control groups for each covariate should be close to zero after adjustment (i.e. the distribution of each covariate is similar).
Typically, a $0.1$ threshold, indicated in the aSMD plot, is used to identify covariates that are not well balanced \textbf{(T4)}.
Users can sort the covariates using the `Sort' button in order to identify covariates that require further analysis.
The toggle buttons above the aSMD plot can be used to switch to a details view, where the distributions of each covariate is visualized individually on a separate plot (Fig. \ref{fig:ceval} \textcircled{3} \& \textcircled{4}).
In order to emphasize covariates that are not well balanced, only those with an adjusted standardized mean difference greater than $0.1$ will be included in the details view by default. 
Users can then manually customize which covariates are shown using the `Show/Hide covariates' dialog (Fig. \ref{fig:ceval} \textcircled{5}). For each covariate, we visualize the control group distribution above and the treatment group distribution below.
To maintain consistency with the aSMD plot, the unadjusted distributions are depicted using a black outline, while the adjusted distributions are filled in.
We also add a dashed line to indicate the mean values for each distribution, with the adjusted means indicated with an additional black dot along the axis.
This supports the rapid comparison of adjusted means since it is necessary to ensure that the adjusted treatment and control distributions are as similar as possible.
A large distance between the treatment and control means would indicate to users that the groups are unbalanced for this particular covariate \textbf{(T4)}.

\textbf{Supporting Different Causal Inference Methods.} From our interviews with causal inference experts, we found that data analysts would approach cohort construction using different methods, such as matching or IPW (see \ref{cohort_construction}).
When using methods such as matching, data analysts would typically take the observational data (unadjusted cohort) and select samples to form balanced treatment and control groups (adjusted cohort).
When using IPW, however, only the unadjusted cohort is used, and each sample in the unadjusted cohort is weighted by the inverse of its propensity score to create a pseudo-population where the treatment and control groups are balanced.
To account for different methods, the CohortEvaluator module can initialized with the \argument{unadjustedCohort} argument, and the optional \argument{adjustedCohort} argument \textbf{(T6)}.
If only the \argument{unadjustedCohort} is provided, the adjusted aSMD values in the aSMD plot will be calculated using the inverse propensity scores to weight the unadjusted aSMD values.

\subsubsection{Usage Scenario}

In the prior step, we used the DAG module to model the causal structure of the data and identify the relevant covariates (confounders and prognostics).
We now adjust for these covariates through matching to obtain a cohort to use for treatment effect estimation.
In this example, covariate matching results in an initial cohort of 328 students -- 192 in the treatment group (frequent absences), and 189 in the control group (few or no absences).
We then pass this selected cohort to the CohortEvaluator module to ensure that the cohort has comparable treatment and control groups with respect to the identified covariates.
Since we are using the matching approach, we pass the original data set to the CohortEvaluator module using the \argument{unadjustedCohort} argument, while the matched cohort is provided as the \argument{adjustedCohort} argument.


\begin{figure}[h]
  \centering
  \includegraphics[width=\linewidth]{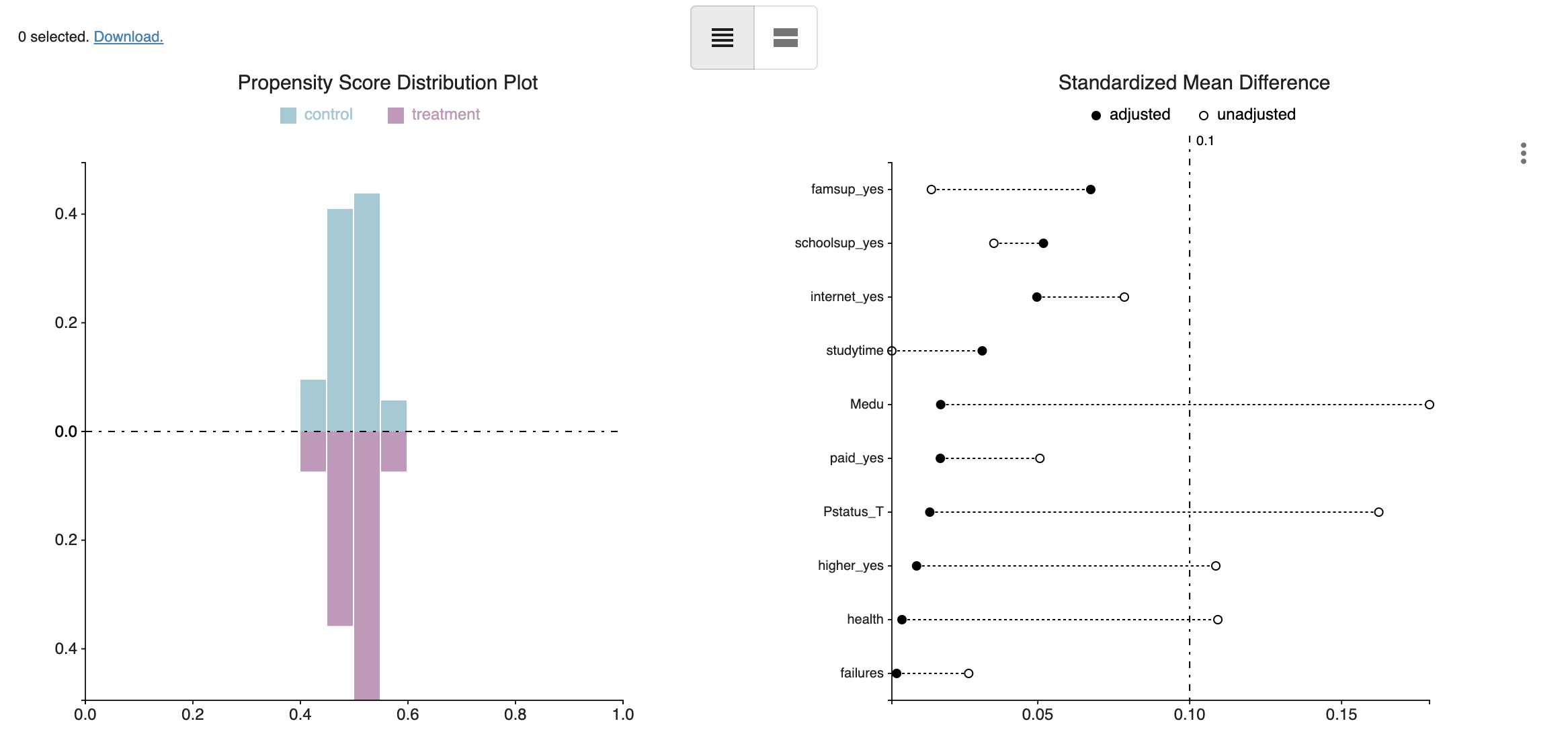}
  \caption{The cohort of selected students after refinement. Compared to Figure \ref{fig:ceval}, we can see from the aSMD plot that each covariate is much better balanced, and the adjusted aSMD of all covariates are now below 0.1 (black points).}
  \label{fig:student_ceval}
  \Description{The CohortEvaluator module visualizing a cohort of selected students after refinement. On the left are two histograms mirrored along the x-axis. The top histogram depicts the propensity score distribution for the control group, and the bottom histogram depicts the propensity score distribution for the treatment group. To the right of the propensity score histograms is the aSMD plot. Covariates are listed down the y-axis. For each covariate, there are two points on the graph connected by a horizontal line. One point is filled while the other only has an outline. The filled point indicates aSMD value after adjustment. The outlined point indicates aSMD value before adjustment.}
\end{figure}

From the CohortEvaluator module, we see that the cohorts are not fully compatible for a causal analysis (Fig. \ref{fig:ceval}).
The propensity scores for the treatment and control groups have different distributions on the left tail, which suggests that there is a positivity violation. The standardized mean difference plot also indicates that the adjusted aSMD of the \textit{internet=yes}, \textit{Medu} and \textit{health} variables are greater than 0.1\footnote{\textit{internet=yes} refers to whether the student has internet at home, \textit{Medu} is mother's education, and \textit{health} is student's health status.}.
In fact, the adjusted aSMD for \textit{internet=yes} is greater than its unadjusted aSMD, which suggests that the matched cohort increased the difference between the treatment and control groups for the \textit{internet=yes} variable.
Taken together, the visualizations in the CohortEvaluator module suggest that the treatment and control groups for this cohort are not sufficiently similar (i.e. do not overlap), which may lead to biases in treatment effect estimation.

We return to the matching process to use more stringent matching parameters.
After this refinement, we obtain a smaller cohort with 200 samples -- 95 in the treatment group (frequent absences), and 105 in the control group (few or no absences).
This cohort is passed to the CohortEvaluator module (Fig. \ref{fig:student_ceval}).
We can see that the propensity score distributions for the treatment and control groups are now more similar.
This is further supported by the standardized mean difference plot, which indicates that the adjusted aSMD for all variables are below 0.1.
This is a well-balanced cohort with no positivity violations, which we can now use in subsequent treatment effect estimation.

\subsection{Treatment Effect Explorer}
The goal of causal inference is often to estimate the average treatment effect (ATE) of a particular treatment on the outcome of interest.
While the ATE is calculated for the entire cohort, it can be useful to compare the effect between different subgroups as well.
If the average effect differs between subgroups (e.g., for males and females), we say that there is a heterogeneous treatment effect.

\begin{figure*}[h]
  \centering
  \includegraphics[width=0.9\textwidth]{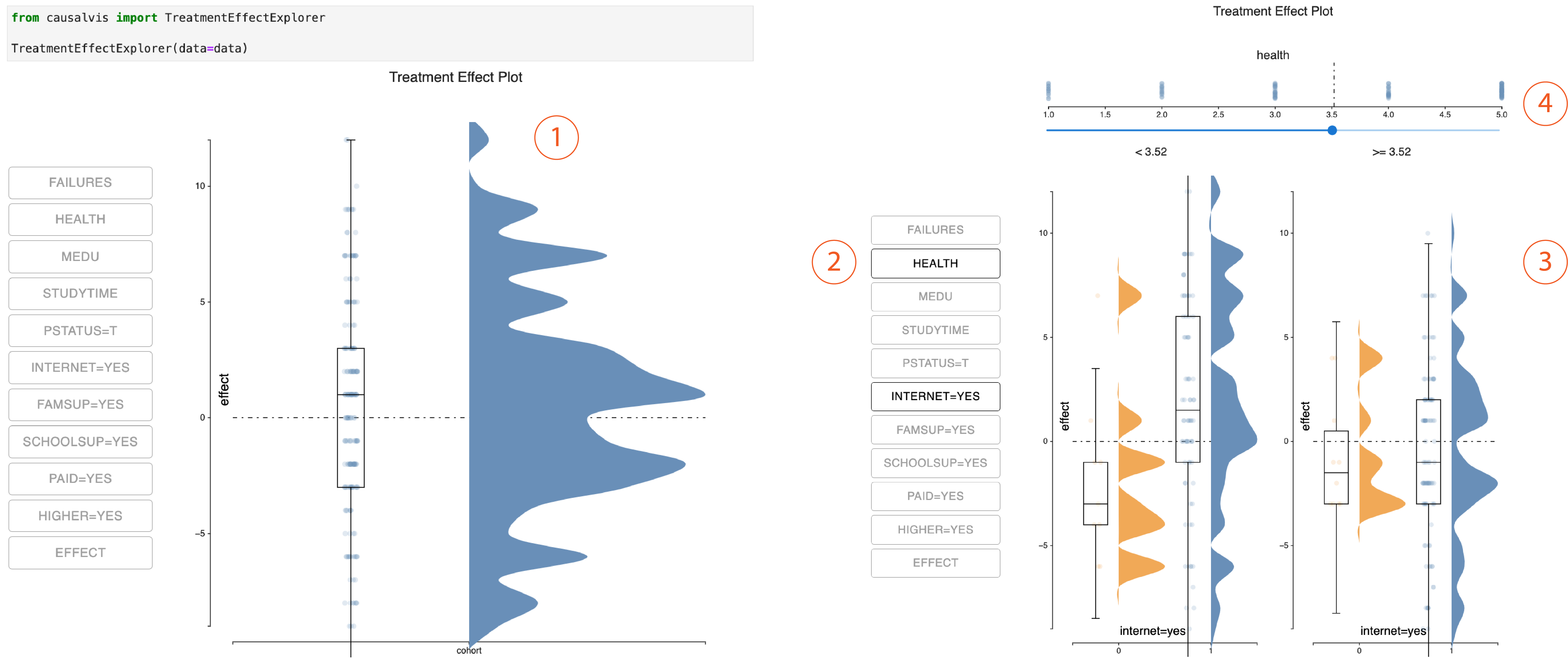}
  \caption{The TreatmentEffectExplorer module. 1) On load, the module shows the distribution of the individual treatment effects of the entire cohort. 2) Users can subgroup the cohort by up to three variables. 3) The visualization will be faceted if two or more variables are selected. 4) If a non-binary variable is selected, the threshold for sub-grouping can be adjusted with a slider.}
  \label{fig:tee}
  \Description{Two screenshots of the TreatmentEffectExplorer module laid out side by side. Parts of the screenshot are labeled from 1-4.}
\end{figure*}

Identifying heterogeneity can result in more precise conclusions \textbf{(T5)}. 
However, to the best of our knowledge, there are no visualization packages that have been developed specifically for treatment effect exploration.
From our formative interviews, we found that causal inference analysts currently make use of general purpose visualization authoring tools such as matplotlib\footnote{https://matplotlib.org/stable/index.html} or seaborn\footnote{https://seaborn.pydata.org/}.
These visualizations are often static and created \textit{ad hoc} for each study, making the process too time-consuming for in-depth subgroup exploration.
The TreatmentEffectExplorer module (Fig. \ref{fig:tee}) is thus designed to visualize individual treatment effects conditioned on different variables, with the goal of helping users identify when trends in treatment effects change across different subgroups (see \ref{treatment_effect_exploration}).
Note that this module can only be used with certain causal inference methods, such as matching, where it is possible to calculate individual treatment effects \textbf{(T6)}.
It is unsuitable for methods such as IPW where only the ATE for the entire cohort can be obtained.

\textbf{Creating and Exploring Subgroups.} On load, the TreatmentEffectExplorer module defaults to a single visualization of the distribution of individual treatment effects for the entire cohort (Fig. \ref{fig:tee} \textcircled{1}). The visualization uses a raincloud plot \cite{allen2019raincloud}, which has been found to convey statistical summaries about the data distribution with minimal distortion or misinformation.
To the left of this visualization is a list of variables available in the data set (Fig. \ref{fig:tee} \textcircled{2}). Users can click on a particular variable name to facet the visualization by this variable.
Up to three variables can be selected this way.
The first selected variable will be visualized along the $x$ axis, while subsequent variables will be used to create small multiples in a matrix layout.
If the first variable is binary, the cohort is automatically divided into two groups based on that variable, and raincloud plots will be used to visualize the distribution for each group separately (Fig. \ref{fig:tee} \textcircled{3}). If the first variable is a continuous variable, all individual treatment effect values will be visualized using a scatterplot.
For the subsequent variables used to create small multiples of the visualization, if the selected variable is binary, the cohort will simply be divided based on that variable.
However, if a faceting variable is continuous, the TreatmentEffectExplorer module first calculates the variable mean, which is then used to divide the cohort into two sub-populations.
Note that the variable mean is only a heuristic used to perform a default grouping of the cohort, and users can customize this threshold value using a corresponding slider bar (Fig. \ref{fig:tee} \textcircled{4}). Additionally, we also visualize the distribution of the continuous variable in a beeswarm plot directly next to the slider to help users identify natural sub-populations in the data where a more appropriate division should be made.

\textbf{Identifying Heterogeneous Treatment Effects.} All faceted plots in the TreatmentEffectExplorer module share the same $x$ and $y$ axes ranges to support easy comparison across the different visualizations. 
This would also help users identify when the treatment effect for one sub-population is significantly different from others \textbf{(T5)}.
We also add a dashed line to the visualizations when the domain of the $y$ axis includes $0$ in order to highlight when sub-populations have opposite treatment effects, which can be an indication of Simpson's paradox.



\subsubsection{Usage Scenario}

In the previous step, we selected a cohort from the data set consisting of a treatment and a control group.
Since we used the matching method, we can calculate individual treatment effects for each treatment-control pair in the cohort as the difference in \textit{G1} grade if a student had frequent absences in a year (\textit{absences} = 1) compared to if they had few or no absences (\textit{absences} = 0).
This then allows us to explore trends in individual treatment effects across the selected cohort.
The individual treatment effect values are passed to the TreatmentEffectExplorer module to explore subgroups and identify heterogeneous treatment effects.


Using the TreatmentEffectExplorer module, we choose to group the cohort based on internet access (\textit{internet=yes}) and student health (\textit{health}) (see Fig. \ref{fig:tee}, \textcircled{3}). The visualizations are now faceted by student health -- the students with poorer health (\textit{health} < 3.5) are visualized in the facet on the left, and students with better health (\textit{health} > 3.5) are visualized in the facet on the right.
Across the four subgroups, we can see immediately that students with poor health who had no internet access at home had a clear the decrease in grades caused by frequent absences.
Comparatively, this effect of absences was less pronounced for the other subgroups, which all have distributions around 0.
This finding may be interesting for parents and educators, and prompt follow up studies into how frequent absences affect the performance of students with health conditions.

\subsection{Version History}
\begin{figure*}[h]
  \centering
  \includegraphics[width=0.9\textwidth]{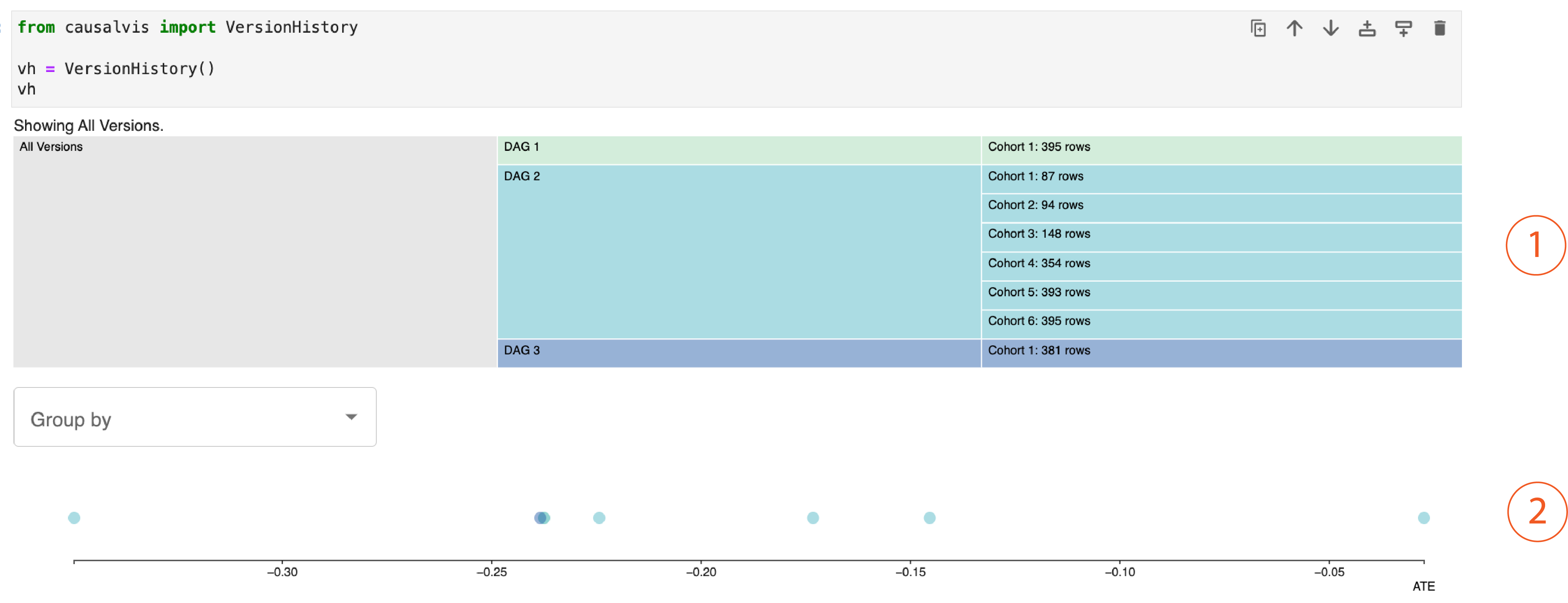}
  \caption{The VersionHistory module. 1) An icicle plot visualizing three versions of the DAG and the different cohorts created to test the causal model in each DAG version. 2) The estimated ATE for each cohort. We see that the estimated values are mostly clustered around the range $-0.25$ to $-0.10$. However, there are two outlier cohorts with very large and very small ATEs.}
  \label{fig:vh}
  \Description{A screenshot of the VersionHistory module. The screenshot has two visualizations. The visualization above is labeled 1. The visualization below is labeled 2.}
\end{figure*}

Causal inference is a highly iterative process where data analysts often have to test different causal structures and construct cohorts that may result in different estimations of ATE.
Keeping track of DAGs and cohorts is thus a crucial task to help users recall their analytic provenance \textbf{(T7)}.
However, to the best of our knowledge, there are no provenance tracking tools that have been developed for causal inference.
For this purpose, the VersionHistory module is designed to store and visualize different DAGs and cohorts such that users can view their causal inference analytic history, as well as restore previous versions when necessary (Fig. \ref{fig:vh}).

\textbf{Tracking Provenance with the VersionHistory module.} The VersionHistory module is initialized simply using \function{VersionHistory()} without any arguments. This creates an empty icicle plot with no DAGs or cohorts.
As data analysts work through the causal inference process, they can track their analytic provenance by calling the \function{.addVersion(}\argument{(DAG, cohort, ATE)}\function{)} function \textbf{(T7)}.
This function accepts a tuple that includes information about the DAG, the cohort selected, and the estimated ATE for this particular iteration of the causal inference workflow \textbf{(T6)}.
The VersionHistory module visualizes DAG and cohort versions using an icicle plot (Fig. \ref{fig:vh} \textcircled{1}). Below the icicle plot is a dot plot that visualizes the ATE corresponding to each cohort. Hovering over a dot will reveal a tool-tip with the DAG and cohort version (Fig. \ref{fig:tee} \textcircled{2}). The visualizations in the VersionHistory module automatically update each time a new version is added, and the user need not run the component again.

\textbf{Saving and Sharing Versions.} At the end of the analysis, all versions can be downloaded using the \function{.saveVersions()} function, which saves all DAGs and cohorts as a \textit{.json} file.
This allows users to easily restore an earlier version of their analysis, as well as share their analytic provenance with collaborators when needed \textbf{(T1)}.

\section{Expert Evaluation}
We evaluate the Causalvis package in a qualitative study with eleven experts to obtain feedback for the visualization modules, and \nobreak validate that the visualizations support the causal inference analysis tasks identified (Section \ref{DG}).

In our future work, we plan to also evaluate Causalvis through a more in-depth deployment study. Working with causal inference experts, we will look at how Causalvis would be used in real-world analyses within specific application domains.
This long-term deployment will help us better understand how Causalvis fits into current user practices.
For the scope of this paper, we rely on participant expertise to evaluate the usability of the Causalvis modules and surface potential issues before real-world deployment. In the following sections, we describe the study design and results.

\subsection{Participants}
We recruited eleven causal inference experts (P1-11, 7 males/4 females) to participate in the evaluation study, five of whom were also participants in the formative interviews (see Section \ref{formative}).
Before each study session, participants were asked to self-report on a scale of 1 (no experience) to 5 (I consider myself an expert) their proficiency in Python ($\mu$=4, $\sigma$=0.775), JupyterLab ($\mu$=3.45, $\sigma$=1.21), causal inference ($\mu$=3.64, $\sigma$=1.12), creating/using DAGs ($\mu$=3.09, $\sigma$=1.04), and data visualization ($\mu$=3.36, $\sigma$=0.674).
All participants are data scientists or performed data analysis as part of their job.
Their experience working with causal inference projects ranged from 3 months to 4 years, with the exception of one participant who has not formally worked on causal inference projects, but who has studied it in his own time and is planning to apply it to subsequent projects.
Of the eleven participants, seven are researchers who have built tools to support causal inference analysis or worked on causal inference research projects in healthcare fields.
Two participants are consultants who work with clients in the government to apply causal inference to policy-related decision making.
The other two participants are doctoral students who use causal inference as part of their doctoral studies.


\subsection{Tasks} \label{evaluation-tasks}
We prepared 4 notebooks in JupyterLab demonstrating each of the 4 modules in the Causalvis package.
In the notebooks, we presented a condensed version of the usage scenario included in this paper (see \ref{usage-scenario}) using the same UCI Student Performance data set \cite{cortez2008using}.
Each notebook included short explanations and examples of how the module would be used in analysis.
Where necessary, we also showed how the module would be used in conjunction with existing Python libraries \textbf{(T8)}.
During the study, we asked participants to work through the notebooks and complete guided tasks such as initializing the DAG with a custom list of variables or brushing over the propensity plot to select samples in the CohortEvaluator module.
These tasks were designed to familiarize participants with the module features.
Once participants completed a notebook, we asked for feedback about the module and improvements they would make.
The sessions were conducted remotely over a video conferencing service, Webex. 
Each session took one to two hours.

\subsection{Results}
Overall, participants gave positive feedback for the Causalvis package.
In the following sections, we highlight when participants found Causalvis to effectively address their analysis tasks (T1-8), and discuss additional feedback and suggestions made for each module.

\subsubsection{DAG}


\textbf{The DAG module supported communication and collaboration (T1). }
Participants found that the option to save and share the DAGs as images was helpful for communicating the causal relationships in the data.
As P4 said, when working on publications of her causal inference work, \textit{``having a good visualization was always pretty important.''}
Similarly, P7 also commented that the visual representation of causal relationships in the DAG was an important means of communicating with audiences who were not data scientists.
Though data scientists would need to know the details about the causal structure and the causal inference process, an external collaborator or domain expert \textit{``only wants to know the relationship between the covariates and see the outcome.''}
In addition to the effectiveness of the visualization itself, participants also felt that the DAG module was an improvement on existing tools.
P9, for instance, described working on publications where he had to manually create DAGs in Powerpoint and Photoshop, concluding that \textit{``this would have been much nicer.''}

\textbf{Automatically identifying different variable types in the DAG module was helpful for participants (T3). } 
During the study, many participants found it useful that the DAG module would dynamically highlight the important variable types once the treatment and outcomes were selected.
As P1 described, \textit{``it's nice that it will immediately color everything so that it's also visibly clear which nodes have which context.''}
Overall, participants felt that this feature was helpful for users who were less familiar with causal inference and \textit{``didn't know the confounders, colliders, and mediators''} (P9), as well as for data analysts who might discover \textit{``something we hadn't realized is a confounder or a mediator somewhere along the graph''} (P4).
Furthermore, participants also liked the use of color encodings in the DAG to highlight each variable type.
By providing visual feedback that made the variable types explicit and quickly identifiable, the DAG module was able to better support the task of selecting the variables that need to be adjusted for.
As P10 described, it \textit{``takes some time to visually parse a graph''}, so \textit{``this automatic annotation is very useful''}.
However, while many participants appreciated this feature, some also cautioned against overreliance on automatic variable identification.
P9 referred to this as \textit{``a double-edged sword''}, and expressed concerns that variables could be wrongly highlighted because someone \textit{``could create a DAG where it labels them incorrectly.''}
This can then introduce errors into the subsequent causal inference analysis if the wrong variables are adjusted for.



\textbf{Participants liked being able to quickly and interactively iterate on the DAG module.}
Compared to the current techniques or tools that are used to create DAGs, many participants commented that they preferred the interactive features provided by the DAG module.
P1, for instance, said that \textit{``it's nice that it gives you the option to edit it manually, and you don't have to write it all in some yaml or something like that.''}
Similarly, P2 had prior experience learning about DAGs from R, but commented that the visualizations had not been interactive, which \textit{``would've been very useful.''}
During the discussion, P9 elaborated further on his preference for interactivity.
Previously, he had created DAGs using tools such as Powerpoint and Photoshop, but felt that \textit{``iterating on something is what makes it really annoying.''}
In comparison, when using the Causalvis DAG module, he appreciated \textit{``being able to quickly iterate and add in and remove nodes and relationships,''} which was made possible through the interactive interface.

\textbf{Participants wanted more annotation features in the DAG module.}
A frequent comment from participants was the suggestion to add annotations to edges between nodes.
For data analysts, this feature would help them gain a better understanding of the data set.
As P1 said, it would be helpful to add the correlation coefficient to the edges because \textit{``it's not just the structure of the causal relationships, but also some numerical estimation of how things relate between one another.''}
This information would help her identify whether a confounder is highly correlated with both treatment and outcome, and would thus need to be prioritized during adjustment.
In contrast, other participants wanted to add annotations to explain relationships or communicate with collaborators.
For example, P9 wanted to \textit{``add information to describe these pathways beyond just an arrow.''}
Similarly, P11 would typically meet with collaborators only every one or two weeks, and wanted to highlight questionable links and indicate parts of the DAG that need to be \textit{``reviewed by the subject matter expert.''}
Ultimately, for data scientists who have to work with domain experts frequently over the course of the causal inference process, annotations can be a helpful means of supporting collaboration.
This is best summarized by P10, who said that
\textit{``[annotations] are not about causal inference. These are just about communication. But in a client setting, communication is important, so that, I think, is useful.''}

\subsubsection{Cohort Evaluator}

\textbf{The detailed covariates view in the CohortEvaluator module provided additional insight into the data distribution in cases of covariate balance and positivity violations (T4).}
Many participants in our study appreciated that they could see the distributions of each covariate visualized separately in the details view (Fig. \ref{fig:ceval} \textcircled{4}), and felt that it a useful addition to the propensity score and the aSMD plots that are more commonly used in causal inference.
Participants described the benefit of the additional plot in the following ways.
As P1 explained, \textit{``when you have, for example, the aSMD plot, you only see the average. I think it's better to see the entire distribution, it gives you much more information.''}
P2 further elaborated on his prior experiences, commenting that \textit{``Whenever I've done propensity score matching, if I want to look at exploratory plots, I would just look at [the covariates] one by one.''}
In comparison, when using the CohortEvaluator module, P2 said that \textit{``not having to write individual code to obtain each of these, I think that's nice.''}

\textbf{Participants wanted feedback from the CohortEvaluator module after selecting instances from the propensity score plot.}
During the expert evaluation study, participants found it useful that they could brush over the propensity score plot to obtain the samples that were not well balanced.
P5, for example, said that \textit{``I really like scrolling over the [propensity score plot] and having a look at which samples they were. It is pretty amazing.''}
However, many participants wanted more visual feedback from the module after making the selection, such as visualizing the covariate values of the selected samples separately.
P5 said that \textit{``if we could have some visualization about the selection that is unbalanced, I think it would be really interesting. A straight, fast visualization about why your data isn't being sampled in those cases.''}
Similarly, P3 wanted to compare the covariates of the selected samples with the entire cohort because \textit{``it will be quite interesting to see the contrast in how the distribution looks.''}
In addition to visualization feedback, some experts in the study also wanted the Jupyter notebook to provide a more detailed explanation about the visualizations and how to interact with them.
For example, P10 said that even after brushing over the propensity score plot, \textit{``it's still not clear what these people are''} and what covariates differentiate them from the rest of the cohort.
In such cases, he wanted the notebook to provide examples about what analysis steps to take next -- \textit{``It doesn't have to change the design, it doesn't change your package. Just add some use cases in here, some action you can take.''}
P11 made a similar suggestion, saying that the notebook examples should include \textit{``more guidance on what packages you recommend''} to characterize and exclude samples after they have been selected.

\subsubsection{Treatment Effect Explorer}

\textbf{The interactive visualizations in the TreatmentEffectExplorer module reduced the effort needed for participants to iteratively explore and compare subgroups in the data (T5). }
Many participants in our study found the visualizations to be an improvement from their current approaches.
P1, for instance, said that \textit{``it's nice to visualize everything and not just look at numbers.''}
Similarly, P8 \textit{``really liked the three-variable visualization [because] it's a really hard thing to do.''}
Participants also described the interactive selection of different variables as being \textit{``very intuitive''} (P4) and \textit{``great fun''} (P6).
The interactivity was helpful to explore different subgroups, and \textit{``get these different comparisons between the average treatment effect''} (P4).
It was also particularly useful for participants who 
often worked with domain experts to identify heterogeneous subgroups in the data.
During exploration, domain experts may ask to compare different variables or stratify subgroups by different thresholds.
As P6 described, with current approaches there is often \textit{``this endless repetition of change''}. In comparison, the TreatmentEffectExplorer was \textit{``easy to work with''}.

\textbf{The TreatmentEffectExplorer module supported communication and storytelling about causal inference.}
In addition to analysis tasks, a number of our participants also mentioned that \textit{``it's the job of the data scientist to come up with the story''} (P11).
In particular, consultants need to communicate and make sense of results for clients, customers, and collaborators.
However, this can be challenging with causal inference.
As P10 described, \textit{``most data scientists know that they have to do some storytelling, but in the causal inference setting, I think because the idea is so new, most data scientists don't know how to tell a story with causality because the story is harder to tell.''}
To this end, participants commented that the visualizations in the TreatmentEffectExplorer module were well suited for this purpose.
Compared to downloading the data and creating visualizations manually, P11 liked that \textit{``this tool can have some graphs ready made to support our story... it saves us a lot of time.''}

\textbf{More visualization customization and guidance should be provided in the TreatmentEffectExplorer.}
Although the visualizations and interactions implemented in this module were well received by our participants, not all participants were familiar with the raincloud plot.
P5 found that \textit{``it took me a while to get into the visualization''} and P3 commented that \textit{``it took less than a minute, but it wasn't immediate.''}
To better interpret the visualizations, many participants wanted the option to customize the plots based on their familiarity and expertise.
P8, for example, suggested putting \textit{``some of the data that you visualize, such as the box plot, on a toggle that you can turn off and on''}, while P4 wanted to \textit{``enable or disable the different [violin plots] because usually I don't plot these.''}
Ultimately, participants wanted more instruction on how to interpret the visualizations in the TreatmentEffectExplorer module.
P4 suggested implementing \textit{``an instructional manual before hand so you know what you're looking at''} in the visualization, while P3 wanted \textit{``something to guide you to what you're looking for''}.

\subsubsection{Version History} \label{evaluation_vh}

\textbf{The VersionHistory module was helpful for tracking provenance but should include additional features (T7). }
Many participants mentioned that keeping track of their analysis process was something that they want or need to incorporate into their workflow.
Referring to their causal inference studies, P2 said that \textit{``I absolutely needed [such a tool]. I absolutely need to record how many patients are in my cohort''}, while P8 commented that \textit{``[for] what I'm doing right now, [version control] would be really helpful because I am running a lot of data against a lot of DAGs.''}
Visualizing the different cohorts and associated ATEs also has the added benefit of helping analysts evaluate the robustness of their estimated ATE when using a cross-validation approach.
As P1 explained, \textit{``you would like to see that the average treatment effect is mostly stable, that it resides in some range that is not very large. Then you can say that we're really robust, and there is real generalization in the model instead of overfitting to whatever data we get.''}
However, some participants also suggested that the VersionHistory module needed to keep track of additional information to be completely useful.
In cases where machine learning models are used in causal inference, the module should also record \textit{``machine learning parameters''} (P11) and \textit{``the specifications of the model''} (P6).

\textbf{Participants wanted the VersionHistory module to help them compare the DAGs.}
In addition to keeping track of the analysis process, multiple participants mentioned that they wanted the VersionHistory module to help them compare between different versions of the DAG.
Participants suggested that a visualization would reduce the effort required to make such a comparison, providing an \textit{``easy way to understand what is DAG 1 and what is DAG 2, and how do they differ from one another''} (P1).
Similarly, P2 said that \textit{``having these comparisons, making these comparisons easier to do would be very helpful.''}
More specifically, participants were interested in using visual comparison to identify unique structures in the causal graphs.
P6, for instance, said that he would like to see \textit{``which edges appear in one and not the other''}, while P8 wanted the module to \textit{``emphasize to me the local structure, what the neighbors of the variables actually are, so I can be more sure that it is not a collider in the path form.''}
P10 went a step further, and suggested keeping track of the version history directly within the DAG module itself because \textit{``while I do the editing, if I have some way to look at the history, that will help.''}
He explained that being able to see previous versions would be visual reminder of earlier discussions with collaborators, which can in turn guide decisions about the edits needed.
Ultimately, there was strong consensus among study participants that being able to compare and contrast different versions of the DAG would enhance the VersionHistory module.
This was best expressed by P3, who said that \textit{``presenting would be a first benefit, but comparing would be even better.''}

\section{Discussion}

\add{
In this section, we reflect on how Causalvis contributes to the existing causal inference workflow, and discuss implications for future work.}

\textbf{Supporting rapid iterative hypothesis testing through interactive visualizations.}
The process of causal inference requires data analysts to iterate through each step of the workflow multiple times to explore different causal structures, refine cohorts, and explore heterogeneous treatment effects in different subgroups.
Participants in our evaluation study appreciated that the interactivity of the DAG and TreatmentEffectExplorer modules allowed for more rapid iterations compared to current tools \add{where static visualizations must be edited programmatically in a manual and time-consuming process}.
Causalvis thus better supported tasks such as collaboratively exploring different hypotheses of causal structure \textbf{(T1)} or comparing heterogeneous treatment effects across cohort subgroups \textbf{(T5)}.
However, participants also felt that more could be done for the CohortEvaluator module.
Although users could brush over the propensity score plot to select samples that are imbalanced between the treatment and control groups \textbf{(T4)}, the visualizations in the module did not update in response to this interaction.
Users had to manually exclude the selected samples from the data set themselves and run the module again in order to see the visualization updates.
Many participants in the study found that this process was unintuitive, and expected the visualizations to dynamically update with indications of how the selected samples differed from the entire cohort, which can then provide the guidance needed to adjust their selection and refine the cohort.
Future work developing visualizations for causal inference should thus better consider the need for rapid iteration throughout the workflow, with particular emphasis on cohort construction and refinement tasks.

\textbf{Explaining and communicating causal inference to domain experts and collaborators.} In causal inference, data scientists do not merely complete analysis tasks, they frequently also have to communicate with domain experts, publish results, and make sense of the outcomes for clients and collaborators.
To this end, many participants in our evaluation study liked when the visualizations in Causalvis were effective for both analysis and communication.
Participants found that the saving and sharing features in the DAG module to be an improvement over existing tools for visualizing causal structure models.
They also liked that the TreatmentEffectExplorer module helped them rapidly identify heterogeneous treatment effects as part of the analysis, while also explaining results to collaborators and \textit{``tell[ing] a story with causality''} (P10).
Taken together, this highlighted the need for causal inference visualizations to support both analysis and communication.
Our evaluation study also revealed the potential use of annotations to keep track of discussions with domain experts.
From our formative and evaluation studies, we found that collaborations during the causal inference workflow can be inconsistent.
Data scientists may only meet domain experts every one or two weeks, during which time they must quickly validate changes and make the necessary refinements.
This is particularly important when modeling causal structures, and many participants suggested adding annotation features to the DAG to better support this two-way collaboration.
In future work, researchers can further explore how annotations might be incorporated into the various visualization modules to better support communication and collaboration needs during causal inference.
Existing studies such as \cite{stokes2022striking} and \cite{choudhry2020once} can also inform the design of these annotation features.

\textbf{Evaluating estimation robustness through sensitivity analysis.}
A characteristic of causal inference is the lack of ground-truth data that can be used to evaluate the accuracy of estimated treatment effects.
Instead, causal inference analysts would often perform sensitivity analysis \cite{iooss2017introduction} to check for robustness and generalizability of various analytical choices.
Analysts may thus iterate through the causal inference process multiple times, using different combinations of covariates or testing different cohorts in each iteration \cite{lei2018distribution}.
As mentioned by P1 during the evaluation study, the VersionHistory module potentially supports a sensitivity analysis-like evaluation (see \ref{evaluation_vh}).
For example, if the ATE remains stable across different covariates and cohorts (see \ref{fig:vh} \textcircled{2}), analysts can then be more confident that their estimate is likely to be robust and generalizable.
This combinatorial testing of reasonable analysis decisions shares many similarities with multiverse analysis \cite{simonsohn2020specification, steegen2016increasing}, where an automated, exhaustive search of all combinations of analytic decisions are made to ensure the robustness of results.
Although Causalvis is designed to be highly interactive and does not naturally lend itself to exhaustive combinatorical searches, analysts can still specify and iterate through the most plausible analytic scenarios. 
As such, many visualization strategies that have been developed to evaluate and review the results of multiverse analysis may be similarly effective for visualizing the subset of causal inference alternatives explored.
Most immediately, for example, the VersionHistory can act as an equivalent of a forest plot for contrasting ATEs under different design choices.
In future work, we hope to turn to existing studies such as \cite{liu2020boba} and \cite{kale2019decision} to inform how the module might be extended to help causal inference analysts perform sensitivity analysis and evaluate for estimation robustness.

\begin{figure}[h]
  \centering
  \includegraphics[width=\linewidth]{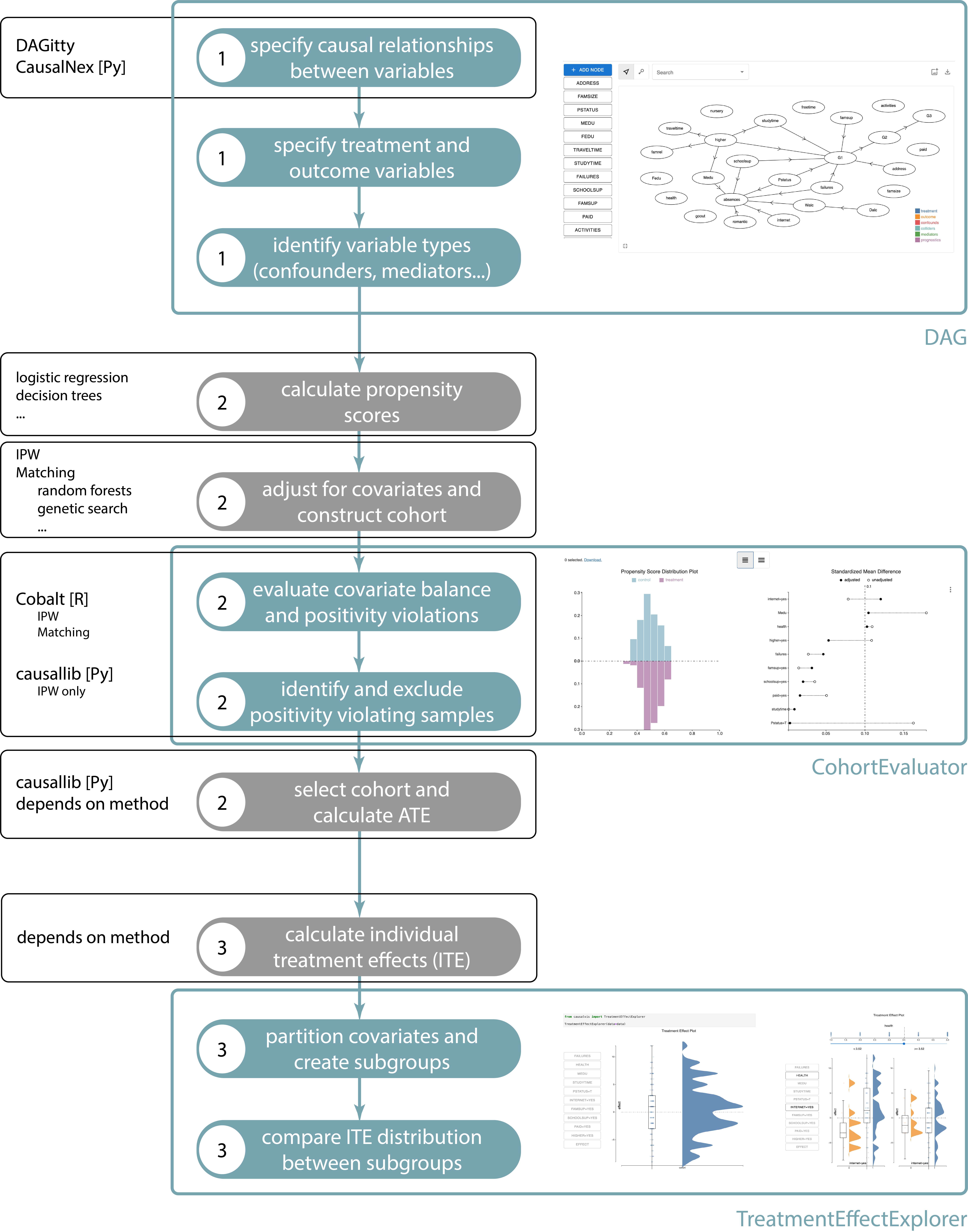}
  \caption{A summary of the causal inference process. Each step in the three-step workflow (1-3) is divided into more granular analysis activities. The analysis activities supported by Causalvis are highlighted in green. Gray boxes indicate other activities supported by existing packages or algorithms. Here, we provide a subset of examples, not an exhaustive list of existing tools. For clarity, we omit arrows indicating iteration between steps.}
  \label{fig:summary_causalvis}
  \Description{11 text labels connected with arrows from top to bottom. The first three labels correspond to step 1, the next 5 label correspond to step 2, the last 3 labels correspond to step 3. Some labels are enclosed in boxes that include lists of libraries or screenshots of a Causalvis module.}
\end{figure}

\section{Lessons Learned from Design Study}
Reflecting on the process of conducting this design study with causal inference experts, we identified some challenges we encountered and the methodological lessons learned.

\bc{
\textbf{Designing visualization packages to complement existing workflows and tasks (T6).}
One key decision we made during the design study was to identify how Causalvis modules should fit into current workflows and complement existing analytic tools instead of interrupting or replacing them.
Doing so required us to understand current causal inference processes and packages.
Working closely with causal inference experts during formative interviews, we refined our understanding of causal inference into the three-step workflow presented in this paper (see Fig. \ref{fig:workflow_diagram}).
We also identified where certain analytic tasks are not well supported by current tools, and made note of the data formats and analysis outputs produced at each step.
This formative study helped guide the development of the Causalvis modules.
For each module, we focused on developing visualizations for tasks that are not well supported by existing tools, while also integrating with the external packages or algorithms analysts have developed (see Fig. \ref{fig:summary_causalvis}).
For example, with the CohortEvaluator module, we learned that analysts computed propensity scores using a variety of different algorithms. We thus decided not to re-implement propensity score calculation in our module, and instead accepted calculated values as input in order to complement the diversity of existing user approaches.
Our paper demonstrates how a visualization package for causal inference can be designed to \textit{complement existing analytic workflows and tools instead of interrupting or replacing them}.
In domains beyond causal inference, there may exist data analysis tasks that require multiple steps and similarly complex workflows.
The workflow discovery process we adopted in this paper may thus be applicable to these other domains as well.
Future research can also investigate comprehensive criteria for making design decisions such that visualization tools optimally complement existing analytic approaches and user tasks.
}



\textbf{Integrating into computational environments (T8).}
Ideating through early designs of Causalvis, we initially considered implementing the visualization modules in an independent visual analytics system outside the interactive computing environment.
However, we ultimately decided against this approach because causal inference experts in our formative study mentioned that they typically worked in the JupyterLab computational environment.
Switching between systems was likely to interrupt users' workflows \cite{elmqvist2011fluid, bederson2004interfaces}, and introducing a standalone application outside the computing environment can be disruptive during highly iterative analyses.
We thus decided to develop Causalvis as visualization modules for use within the JupyterLab computational environment.
To do so, we had to address challenges in both the way the modules were designed, as well as how they were evaluated.
During module design, we incorporated JupyterLab into the earliest prototypes and wireframes of each module, using screenshots of pseudo-code to demonstrate how the visualization modules would fit into the computational environment.
This helped surface more detailed requirements about the inputs and outputs of each module.
For example, after viewing the initial wireframes, an expert in the formative study requested that the DAG module also support networkx graphs as input so that he can easily pass the outputs of the Causalnex package directly into Causalvis.
Similarly, in later evaluation, we created notebooks to demonstrate how the visualizations would be used in JupyterLab, and asked participants to complete tasks that included directly modifying the notebook content (see \ref{evaluation-tasks}).
Through this study design, we identified useful feedback for how each module can be better integrated into JupyterLab.
For example, participants asked questions about data types \textit{(``Why do we convert to a dictionary?'', P11)} and made suggestions about expected outputs \textit{(``I would want the output here to be indices in the data set rather than the data itself'', P3)}.
By conducting the study within the intended usage environment, we were thus able to evaluate how well each module integrated with the computational environment itself, and how the API might be improved.

\textbf{\bc{Working with experts in diverse domains and roles.}}
\bc{In this study, we were fortunate to receive feedback from experts through online meetings on three occasions (as detailed in Section~\ref{sec:study_procedure}). The first session aimed to identify users' workflows for conducting causal inference analysis, the second session was for brainstorming and confirming functionalities based on prototypes and sketches, and the final evaluation aimed to assess the usability of Causalvis. One key lesson we learned was the importance of recruiting a diverse group of participants and understanding user workflows from different perspectives. All of our participants were experts in causal inference, but they had different domains and roles for day-to-day data analysis activities.
Learning from a diverse group of participants allowed us to design for a broad range of use cases.
For example, consultants valued using visualizations to explain the process and ``tell a story" about causal inference to collaborators, while healthcare data analysts were most interested in comparing treatment effects across different subgroups.
We conjecture that we would have had only partial solutions if we did not have access to a diverse group of participants, and we learned that it is important to discover applications from diverse users when designing a visualization package for a specific task.}






\section{Conclusion}
In this paper, we present a design study for Causalvis, a Python package of visualizations to support causal inference.
Working closely with the experts over the course of three months, we first characterized the causal inference workflow and identified related analytic tasks.
We then adopted an iterative design process to develop four visualization modules to support the tasks of causal structure modeling, cohort construction and refinement, and treatment effect exploration.
The results of our evaluation study indicate the importance of designing visualizations to support rapid iteration through each step of the workflow, as well as how causal inference tools should be designed for both analysis and communication.
Finally, we also shared methodological lessons learned from developing componentized visualizations to support a flexible workflow, as well as how visualizations can be designed and evaluated for integration into specific computational environments.
In future work, we aim to further explore how visualizations for causal inference can be designed to support the visual comparison of DAGs and how annotation features can be implemented to facilitate communication between data scientists and collaborators.

\begin{acks}
We would like to express our gratitude to all participants who generously provided their valuable time and shared insights throughout the study.
We also wish to thank the anonymous reviewers for their thoughtful and detailed feedback on this paper, and members of the Georgia
Tech Visualization Lab and researchers from IBM Research for their helpful suggestions, thoughtful comments, and constructive feedback throughout the study.
This project is supported in part by NSF IIS-1750474 and NSF 2247790.
\end{acks}






\balance
\bibliographystyle{ACM-Reference-Format}
\bibliography{main}


\appendix

\renewcommand{\thefigure}{A\arabic{figure}}
\setcounter{figure}{0}

\section{Causal Inference Tools and Packages}

\subsection{DAGitty} \label{app_dagitty}

\begin{figure}[!b]
  \centering
  \includegraphics[width=\linewidth]{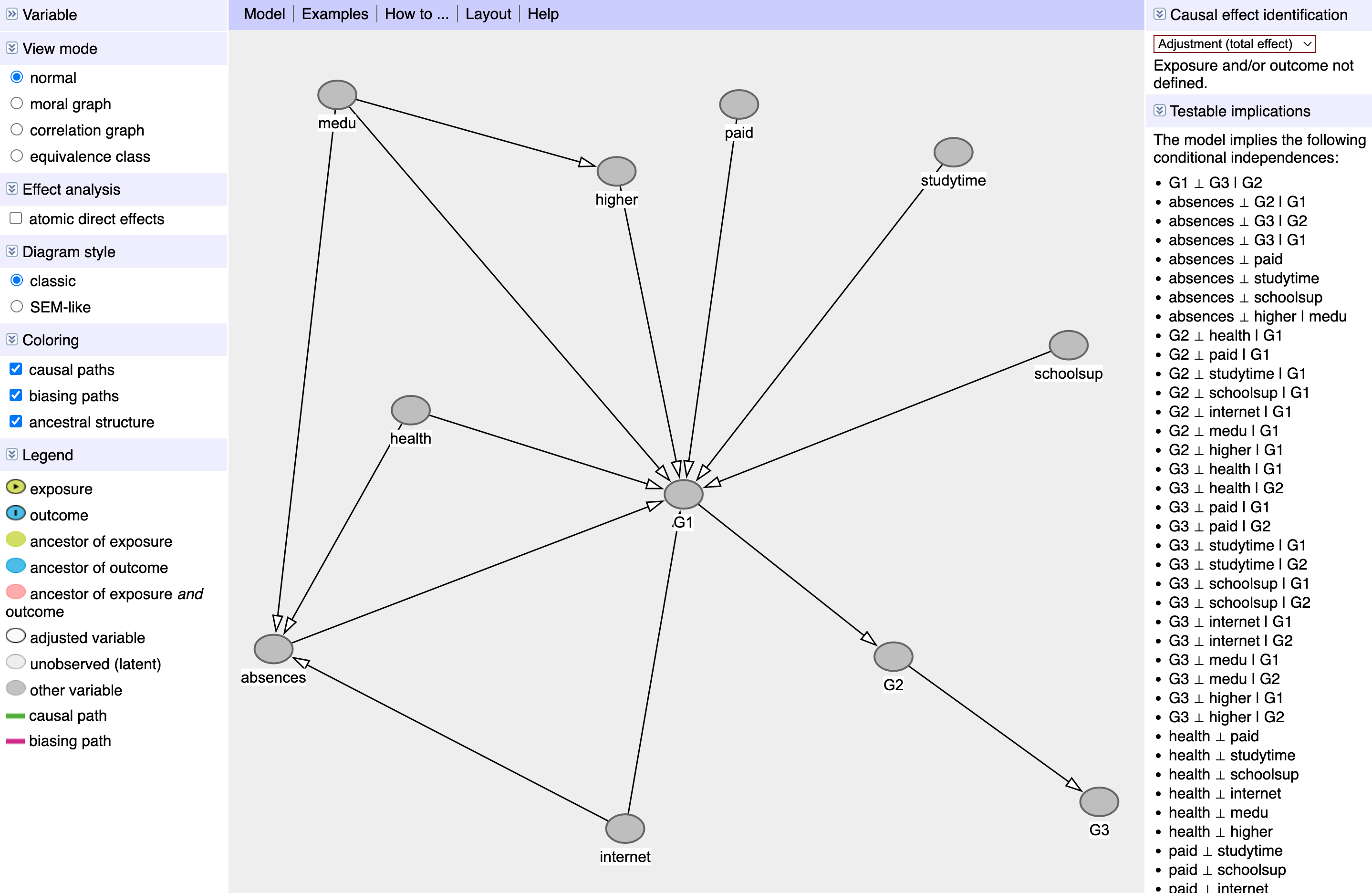}
  \caption{A screenshot of the DAGitty \cite{textor2016robust} application visualizing a DAG of a subset of variables from the UCI machine learning data set.}
  \label{fig:dagitty}
  \Description{The screenshot has a large DAG in the middle. On the left and right are interactive menu options.}
\end{figure}

DAGitty (Fig. \ref{fig:dagitty}) is an application that allows users to interactively specify a causal structure graph.
It is a powerful and comprehensive program, but users are expected to know causal structure terminology (e.g. confounders, conditional independence) and notations, and be adept at making use of these concepts.

\subsection{Causalnex} \label{app_causalnex}

\begin{figure}[!b]
  \centering
  \includegraphics[width=\linewidth]{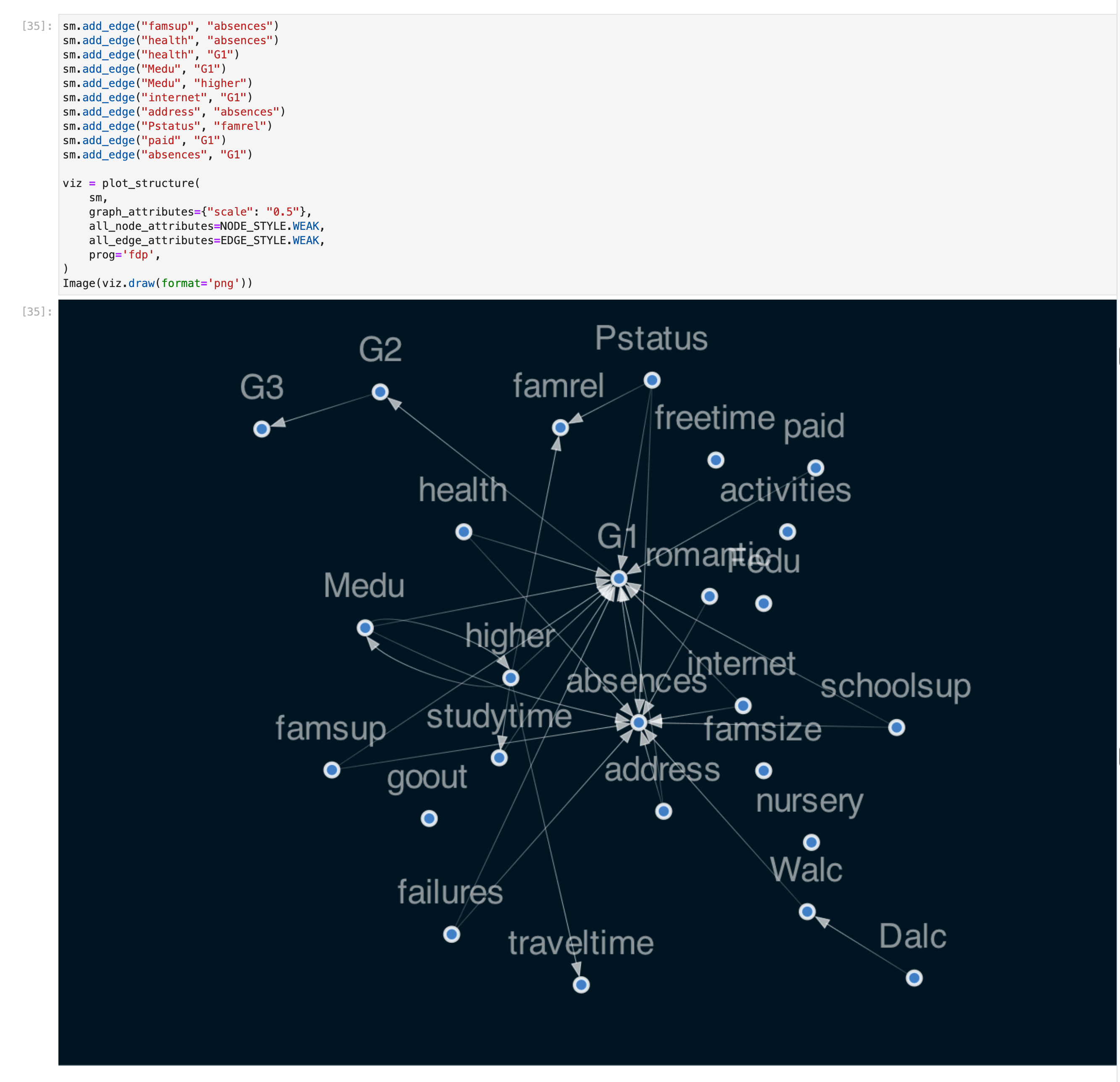}
  \caption{A screenshot of the Causalnex \cite{Beaumont_CausalNex_2021} package visualizing a DAG of the variables in the UCI machine learning data set. The Jupyter notebook cell above shows how an analyst would manually add directed arrows to indicate causal relationships between pairs of variables.}
  \label{fig:causalnex}
  \Description{The screenshot has a large DAG on the bottom. Above the DAG is a Jupyter notebook cell with lines of Python code.}
\end{figure}

Causalnex (Fig. \ref{fig:causalnex}) is a python package that performs causal discovery and allows users to model causal structures by
specifying nodes and links between nodes.
The DAG visualizations in Causalnex are static and must be edited by manually writing code.

\subsection{Causallib} \label{app_causallib}

\begin{figure}[tbh]
  \centering
  \includegraphics[width=\linewidth]{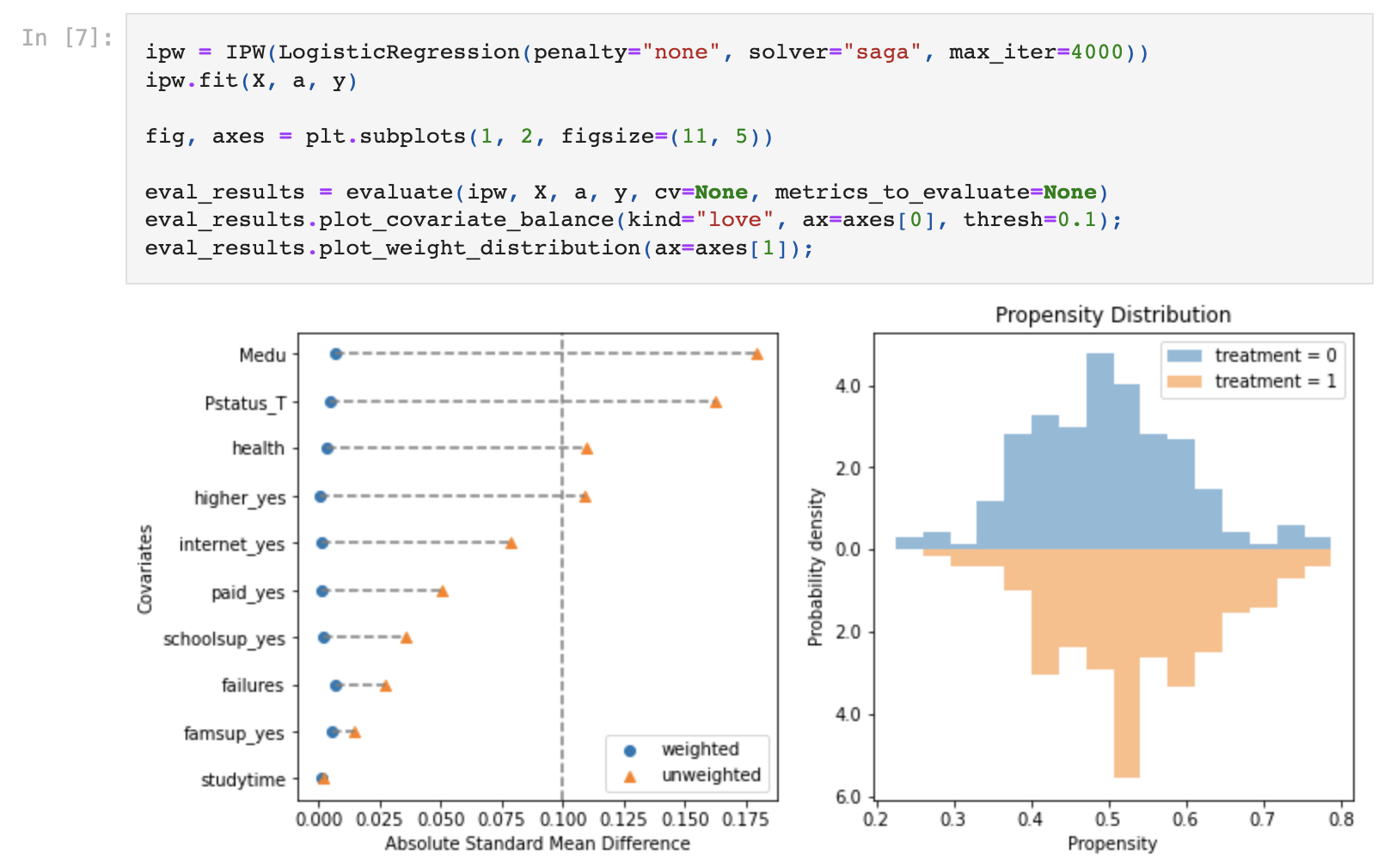}
  \caption{A screenshot of the causallib \cite{causalevaluations} package visualizing the aSMD plot and propensity score distribution for a cohort after covariate adjustment using IPW.}
  \label{fig:causallib}
  \Description{The screenshot has two visualizations on the bottom. The visualization on the left is the aSMD plot with covariates listed down the y-axis. The visualization on the right is the propensity score distribution plot that shows two histograms mirrored along the x-axis. Above the DAG is a Jupyter notebook cell with lines of Python code.}
\end{figure}

Causallib (Fig. \ref{fig:causallib}) is a Python toolkit that implements a range of machine learning models for causal inference methods including the popular IPW and matching methods discussed in this paper.
Visualizations are included as part of the package to help analysts evaluate the machine learning models used.
Closely related to our work are the aSMD plot and propensity score distributions included as part of the IPW methods.
All causallib visualizations are static. 

\subsection{Cobalt} \label{app_cobalt}

\begin{figure}[tbh]
  \centering
  \includegraphics[width=\linewidth]{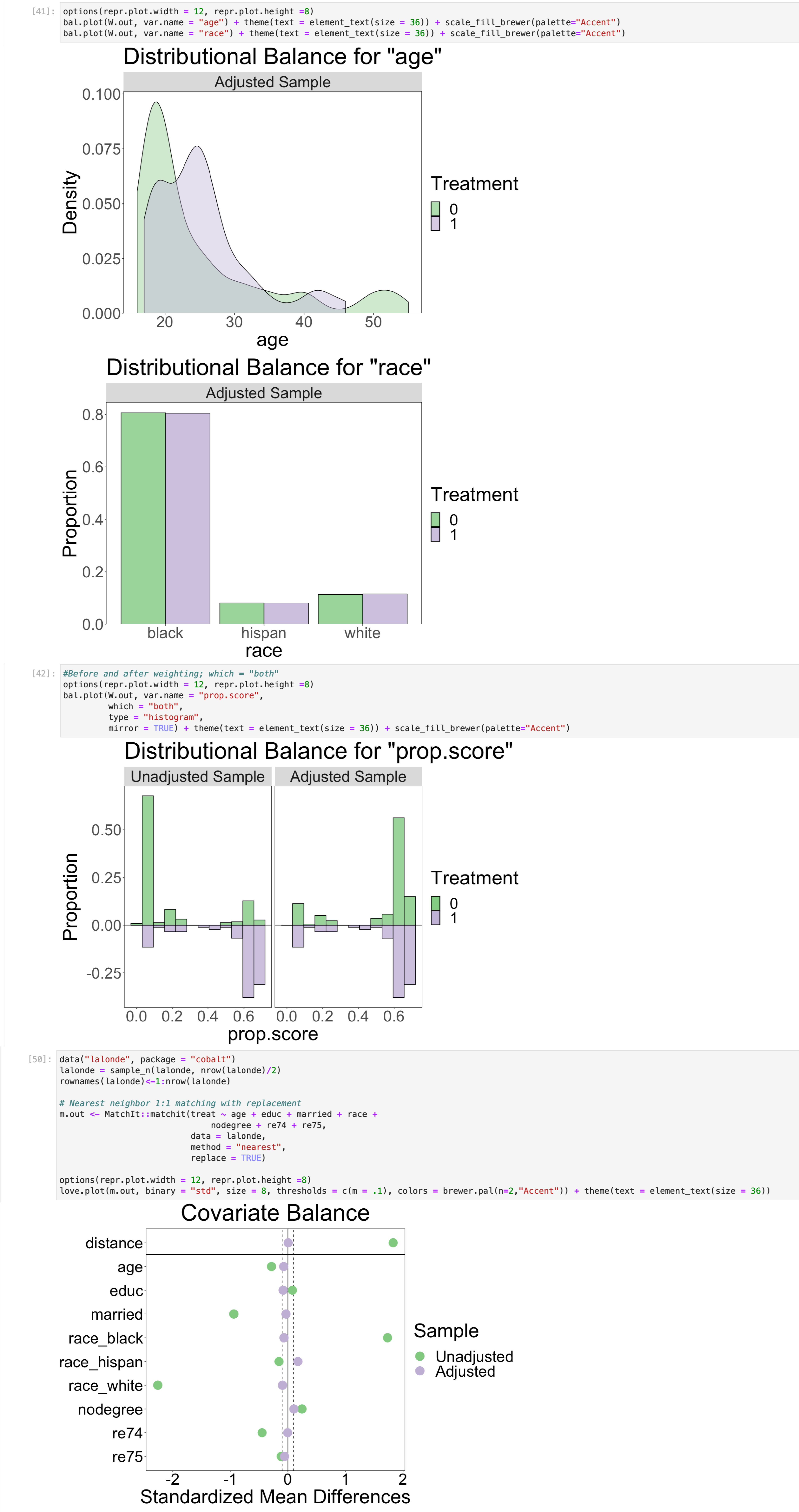}
  \caption{A screenshot of the Cobalt \cite{greifer2020covariate} package visualizing the individual covariate distributions, aSMD plot and propensity score distribution for a cohort after covariate adjustment using Matching.}
  \label{fig:cobalt}
  \Description{The screenshot shows Jupyter notebook cells interleaved with visualizations. The notebook cells include lines of R code.}
\end{figure}

Cobalt (Fig. \ref{fig:cobalt}) is an R package that generates visualizations to help assess the output of causal inference cohorts produced through matching and other methods. Visualizations must be instantiated through separate function calls. All cobalt visualizations are static.

\end{document}